\documentclass[a4paper]{article}
\pdfoutput=1
\usepackage{enumerate}
\usepackage{amsmath}
\usepackage{amssymb}
\usepackage{layout}
\usepackage{ifthen}
\usepackage{url}
\usepackage{pgf}
\usepackage{tikz}
\usepackage[retainorgcmds]{IEEEtrantools}
\usepackage{lscape}
\usepackage{tabularew}
\usetikzlibrary{calc,intersections,through,backgrounds}
\author{Roger Sewell\\
\href{mailto:roger.sewell@cantab.net}{\scriptsize{roger.sewell@cantab.net}}
}
\usepackage{graphicx}
\usepackage{xr}
\usepackage{float}
\usepackage[pdftex,colorlinks=true,linkcolor=blue]{hyperref}

\newboolean{isfalse}
\setboolean{isfalse}{false}

\title{Methods of self-assessment of confidence for secondary school
  maths students, and the benefits or otherwise of using such
  methods\footnote{First version submitted to arxiv.org; RFS version
  1.22.1.1
 .}.
}

\DeclareMathOperator*{\E}{E}

\begin{document}
\maketitle

Acknowledgements: The author is grateful to Colin Foster for inspiring
this work and to Stephen Mack for making the connection. Nonetheless
responsibility for any errors is the author's own.


\begin{center}
\textbf{Abstract}
\end{center}

We first consider the method of scoring students' self-assessment of
confidence (SAC) used by Foster in \cite{FosterCA}, and find that with
it reporting their true confidence is not the optimal strategy for
students. We then identify all continuously differentiable scoring
functions that both drive the student towards the optimal strategy of
truthful reporting of confidence and satisfy an additional axiom
ensuring motivation also to give correct answers to the questions
asked. We discuss the relative merits of some of them, and favour
splitting marks between a signed mark for correctness or not and a
second mark for SAC based on the apparent Shannon information on
whether the answer is correct, as the latter also imparts a useful
life skill, namely avoiding being overconfident.

We then turn to do further Bayesian analysis of the public dataset
associated with \cite{FosterCA}, showing that the effects of
incorporating SAC into teaching vary both by school and by quartile of
ability in class. Finally we speculate on the potential reasons for
this and discuss how future research could identify and avoid some of
the causes.

\tableofcontents

\section{Introduction}

The self-assessment of ones own confidence (SAC) in the correctness of
ones answer to a mathematical question provides the opportunity both
to reflect on how sure one is and on whether more work is needed, and
to receive realistic feedback on ones own accuracy. In
\cite{FosterCA}, Foster investigated the use of a particular method of
SAC in maths teaching and its effect on (standard unmodified)
assessment results by comparing classes using SAC with those not using
it. As in \cite{FosterCA} we will assume the setting of a UK secondary
school teaching students aged 12-18.

The specific method used in \cite{FosterCA} is this: Ask the student,
in addition to providing the answer to the question asked, to also
provide an estimate $q$ of how confident they are that their answer is
correct, giving 10 for totally confident and 0 for totally
unconfident. When the question is marked, a right answer scores $q$
and a wrong answer scores $-q$.

However, it is unclear here whether ``totally unconfident'' was
understood by students to mean ``I'm certain it's wrong'' or ``It's
equally likely to be right as wrong''. Because probability is the
mathematically natural way of expressing such confidence, particularly
in a Bayesian setting, we here instead assume a scale of 0 to 1, with
1 meaning ``I'm certain it's correct'' and 0 meaning ``I'm certain
it's wrong'', and 0.5 meaning ``I think it's equally likely to be
right as wrong''. (For practical application to students who don't
understand non-integer numbers this can of course be rescaled in
practice.)

It is claimed in \cite{FosterCA} that in the long term students cannot
systematically improve their SAC scores by over- or under-stating
their true confidence levels, but this is in fact not true: indeed, a
policy that gives maximum expected score $s$ is to report confidence
$q$ to be 1 (or top of the allowed scale) for any true confidence
value $p$ greater than 0.5 and 0 (or bottom of the allowed scale) for
any $p<0.5$, since $$E(s) = pq + (1-p)(-q),$$ the expectation of the
score on this question, is maximised not by setting $q=p$ but by
setting $$q=\left\{\begin{matrix}1 & (p\geq \frac{1}{2})\\0 &
(p<\frac{1}{2}).\end{matrix}\right.$$ Even so, it is entirely possible
that this was not spotted by the students whose results are analysed
in \cite{FosterCA}, and that it may therefore nonetheless have had the
desired psychological effect.

In the present paper, then, we seek to do two things: first, to
investigate functions for scoring SAC responses with a view to
finding ones with better properties, and second, to further analyse
the data collected in \cite{FosterCA} to look for further clues as to
which students may be affected in which ways, and to suggest future
investigations that might lead to improved learning outcomes for
students. 

\section{Functions for self-assessment of confidence}

\subsection{Introduction}

Indeed, in regard to the first aim, we invite the reader to consider
the following questions. We will denote by $q$ a student's reported
confidence that his answer to a given question is correct, and by $p$
the probability that an answer for which the student reports
confidence $q$ is actually correct\footnote{If you are a frequentist,
you might define $p$ to be the fraction of answers which the student
rates $q$ that are actually correct.}. If the student has correctly
judged his own ability, $p$ will also be equal to his subjective
probability that his answer is correct.

Then, if the score $s(q)$ for a right answer is $f(q)$ and that
for a wrong answer is $f(1-q)$ (or more generally $g(q)$), and $I$
denotes the open interval $(0,1)$:

\begin{enumerate}

\item \label{Q1} Find all continuously differentiable functions $f:(0,1)\to
  \mathbb{R}$ (if any exist) such that $$h(p,q) = \E s(q) = pf(q) +
  (1-p)f(1-q)$$ satisfies, for $J=I$:

\begin{enumerate}

\item \label{C1} $\forall p \in I$, $q\mapsto h(p,q)$ has a single
  strict local maximum on $I$ at $q=p$; 
\item \label{C2} $p\mapsto h(p,p)$ is strictly increasing on $J$.

\end{enumerate}

\item \label{Q2} Are there any more such functions if we instead
  define $J=\left(\frac{1}{2},1\right)$ ?

\item \label{Q3} Find all continuously differentiable functions
  $f,g:(0,1)\to \mathbb{R}$ (if any such pairs exist) such
  that $$h(p,q) = \E s(q) = pf(q) + (1-p)g(q)$$ satisfies conditions
  \ref{C1},\ref{C2} above for $J=I$.

\item \label{Q4} Now suppose that such functions $f$ (or $f,g$) are to
  be used for SAC; discuss the merits and demerits of the various
  options.

\end{enumerate}

Intuitively we want the student's reported confidence $q$ to be
trained to match his actual accuracy $p$, hence the desire to impose
condition \ref{C1}; however we also want to encourage right answers,
and don't want $h$ to be maximised at $h(0,0)$ (``I am sure that my
answer is wrong'', and indeed it is) as one can always easily come up
with a definitely wrong answer (e.g. ``What is 2+2 ?'' ``2+2=Frog''),
hence the desire to impose condition \ref{C2}.

\subsection{Solution to question \ref{Q1}}

Question \ref{Q1} is: Find all continuously differentiable functions
$f:(0,1)\to \mathbb{R}$ (if any exist) such that $$h(p,q) = \E s(q) =
pf(q) + (1-p)f(1-q)$$ satisfies, for $J=I$:

\begin{enumerate}[(a)]

\item $\forall p \in I$, $q\mapsto h(p,q)$ has a single
  strict local maximum on $I$ at $q=p$; 
\item  $p\mapsto h(p,p)$ is strictly increasing on $J$.

\end{enumerate}

\begin{figure}[hpt]
\begin{center}
\includegraphics[scale=0.5]{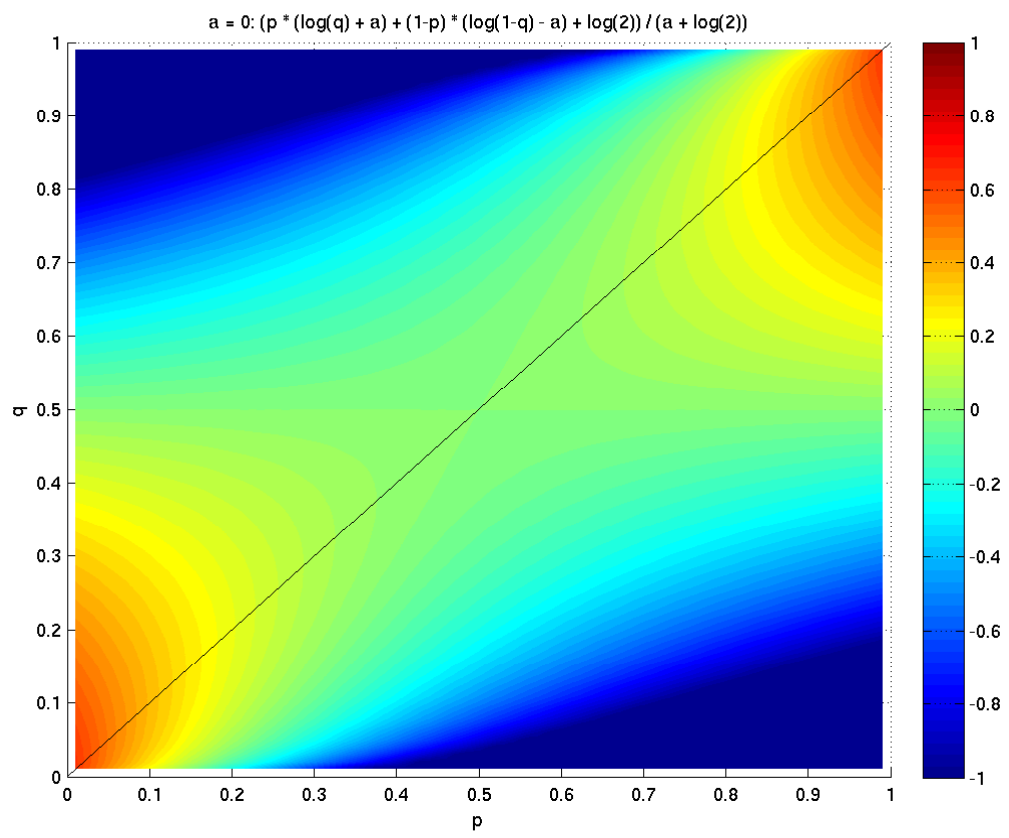}

\caption{Expectation of score using $s(q)=\log_2(2q)$ for a correct
  answer and $s(q)=\log_2(2(1-q))$ for a wrong answer. Note that the
  colour scale has been clipped below at -1, and that in the top left
  and bottom right corners $h$ approaches $-\infty$.
\label{infoplot}
}
\end{center}
\end{figure}

Now, fixing some $p$, if there is a single strict local maximum,
then $$\frac{\partial}{\partial q}h(p,q)=0$$ has a single solution for
$q$ in $I$. Thus $$pf'(p) = (1-p)f'(1-p),$$ which we can achieve for
example by seting $f'(p)=1/p$, which not only achieves an extremum of
$q\mapsto h(p,q)$, but one that is a maximum. Indeed, subject to
multiplying by a constant and adding a constant, we have $$h(p,q) =
\frac{p \log(q) + (1-p)\log(1-q) + \log{2}}{\log(2)},$$ a function
shown in figure \ref{infoplot}.

But now while still satisfying condition \ref{C1} we have freedom to
multiply $f'$ by any continuously differentiable function $m$ such
that for all $q\in I$, $m(q)=m(1-q) > 0$, and indeed any $f$
satisfying condition \ref{C1} must arise in this way.

However, for all $p\in I$ we also have $h(p,p) = h(1-p,1-p)$, so
$p\mapsto h(p,p)$ cannot be strictly increasing, and no such functions
exist that also meet condition \ref{C2}.

\subsection{Solution to question \ref{Q2}}

\begin{figure}[hpt]
\begin{center}
\includegraphics[scale=0.5]{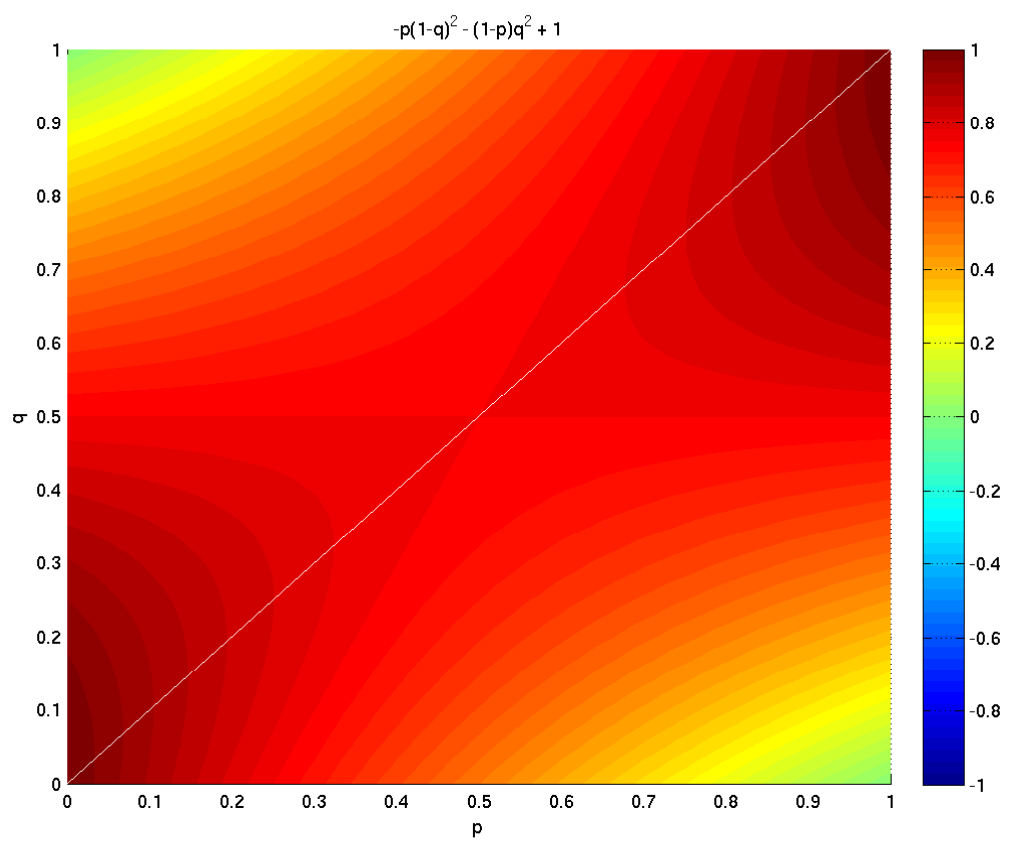}

\caption{Expectation of score using $s(q)=2q-q^2$ for a correct answer
  and $s(q)=1-q^2$ for a wrong answer. 
\label{fig2}
}
\end{center}
\end{figure}

For question \ref{Q2} we relax condition \ref{C2} to restrict $J$ to
be $\left(\frac{1}{2},1\right)$, so the last paragraph of the solution
to question \ref{Q2} no longer applies, and indeed $f(p)=\log(2p)$
satisfies the required conditions, giving e.g. $$h(p,q) = \frac{p
  \log(q) + (1-p)\log(1-q) + \log{2}}{\log(2)},$$ already illustrated
in figure \ref{infoplot}.  Alternatively, taking e.g. $m(q)=2q(1-q)$
we get $$h(p,q) = -p(1-q)^2 - (1-p)q^2 + 1,$$ a function shown in
figure \ref{fig2}.

In regard to satisfying condition \ref{C2} with
$J=\left(\frac{1}{2},1\right)$, we note that for $p\in
J$, 
\begin{IEEEeqnarray*}{rCl} \frac{d}{dp}h(p,p) &=& f(p) - f(1-p) +
  pf'(p) - (1-p)f'(1-p)\\
&=& f(p) - f(1-p) + m(p) - m(1-p)\\
&=& f(p) - f(1-p)\\
&=& \int_{\frac{1}{2}}^p{\frac{m(p)}{p}\,dp} -
  \int_{\frac{1}{2}}^{1-p}{\frac{m(p)}{p}\,dp}\\
&=& \int_{1-p}^p{\frac{m(p)}{p}\,dp}\\
&>& 0
\end{IEEEeqnarray*}
so condition \ref{C2} with $J=\left(\frac{1}{2},1\right)$ is satisfied
by any function constructed thus.

In summary, the set of continuously differentiable functions
satisfying conditions \ref{C1} and \ref{C2} for $J=(\frac{1}{2}, 1)$
are precisely the integrals of pointwise products of the function
$p\mapsto \frac{1}{p}$ with functions $m:I\to \mathbb{R}$ satisfying
$\forall x\in I, m(x) = m(1-x) > 0$.

\subsection{Solution to question \ref{Q3}}

Question \ref{Q3} is: Find all continuously differentiable functions
$f,g:(0,1)\to \mathbb{R}$ (if any such pairs exist) such that $$h(p,q)
= \E s(q) = pf(q) + (1-p)g(q)$$ satisfies conditions \ref{C1},\ref{C2}
above for $J=I$.

Applying condition \ref{C1} we find that for all $p\in I$, $$g'(p) =
-\frac{p}{1-p}f'(p).$$ Let us start then by picking any continuous $f'(p)$
and setting $g'(p)$ as thus constrained. We then have, for each $p\in
I$, an extremum of $q\mapsto h(p,q)$ at $q=p$. Inspection
of $$\frac{\partial}{\partial q}h(p,q) = pf'(q) -
(1-p)\frac{q}{1-q}f'(q)$$ shows that the extremum being unique and a
maximum is equivalent to $f'(q)$ being positive for all $q\in
I$. Checking $$\frac{\partial}{\partial p}h(p,p) = f(p) - g(p)$$ we
note that we also require $f(p) > g(p)$ for all $p\in I$, and as
expected these conditions together then are equivalent to conditions
\ref{C1} and \ref{C2} being satisfied. So long as the $g$ resulting
is bounded above we can achieve the last inequality by simply
adding a constant to $f$.

\begin{figure}[hpt]
\begin{center}
\includegraphics[scale=0.5]{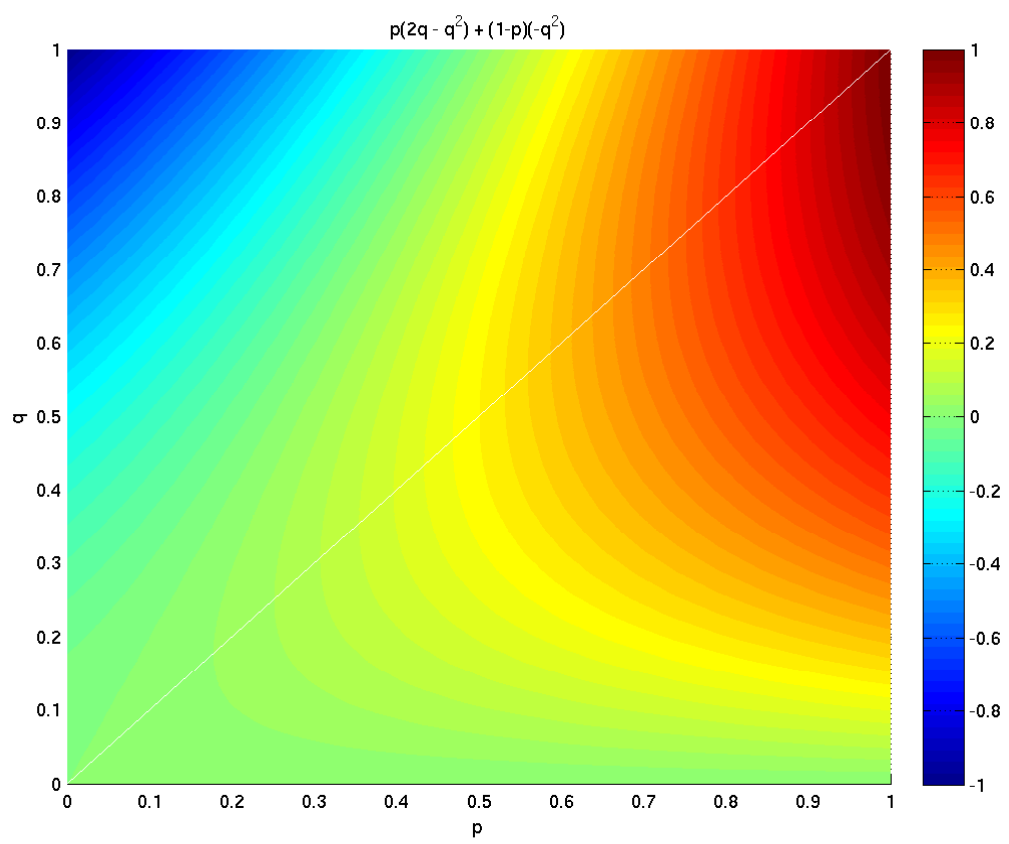}

\caption{Expectation of score using $s(q)=2q-q^2$ for a correct
  answer and $s(q)=-q^2$ for a wrong answer. 
\label{fig3}
}
\end{center}
\end{figure}

\begin{figure}[hpt]
\begin{center}
\includegraphics[scale=0.5]{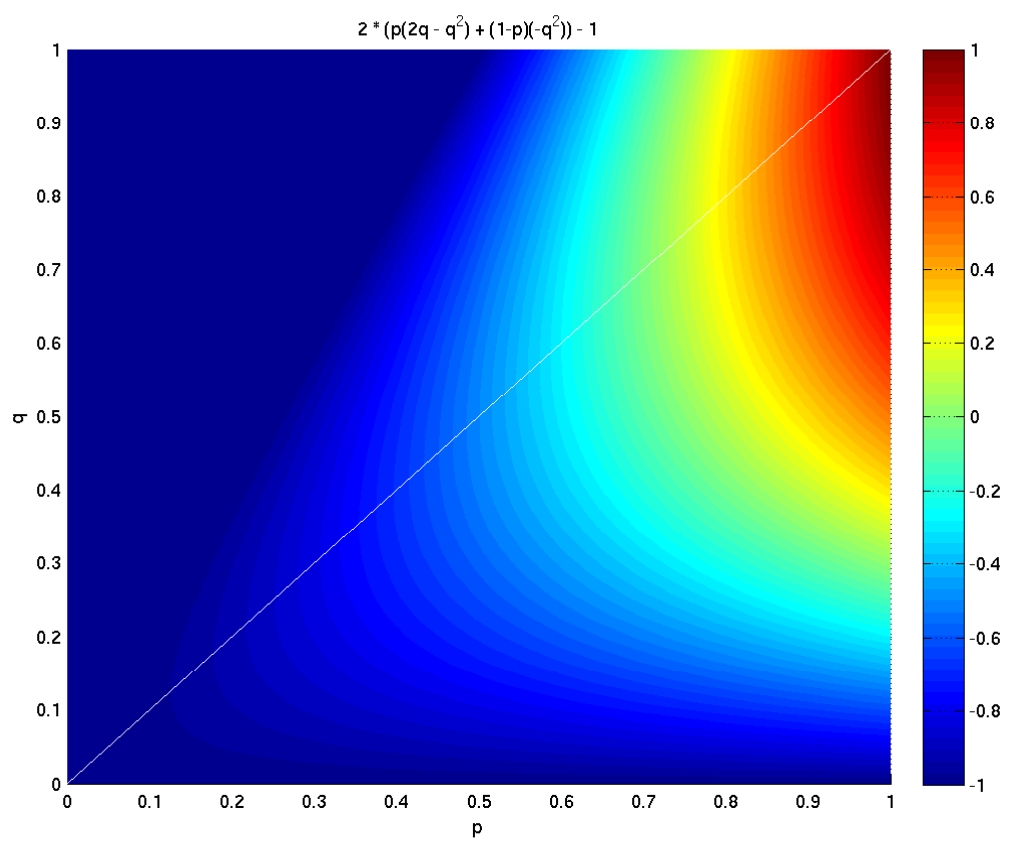}

\caption{Expectation of score using $s(q)=2(2q-q^2)-1$ for a correct
  answer and $s(q)=-2q^2-1$ for a wrong answer. Note that the colour
  scale has been clipped below at -1, and that $h$ approaches -3 in
  the top left corner.
\label{fig4}
}
\end{center}
\end{figure}

As an example we set $$f'(p) = 2(1-p),$$ $$g'(p)= -2p,$$ so
that $$f(p) = 1-(1-p)^2,$$ $$g(p)=-p^2,$$ and thus $$h(p,q) = p(2q -
q^2) + (1-p)(-q^2),$$ a function shown in figure \ref{fig3}, or after
multiplying by 2 and subtracting 1 in figure \ref{fig4}.

In summary the set of continuously differentiable pairs of functions
$(f,g)$ such that $h$ given as in question \ref{Q3} satisfies
conditions \ref{C1} and \ref{C2} is precisely the set of integrals of
functions $f'>0$ and $g'(p)=-\frac{p}{1-p}f'(p)$ such that for all
$p\in I$, $f(p) > g(p)$.

\subsection{Discussion of question \ref{Q4}}

We now discuss the merits and demerits of these various options for
SAC. All achieve the primary goal of training students to correctly
assess their own accuracy by satisfying condition \ref{C1}. However,
using the version of figure \ref{infoplot} or \ref{fig2} alone
completely disregards whether the student has correctly answered the
question, giving him full marks for a wrongly answered question that
he reports as being definitely wrong.

\begin{figure}[hpt]
\begin{center}
\includegraphics[scale=0.5]{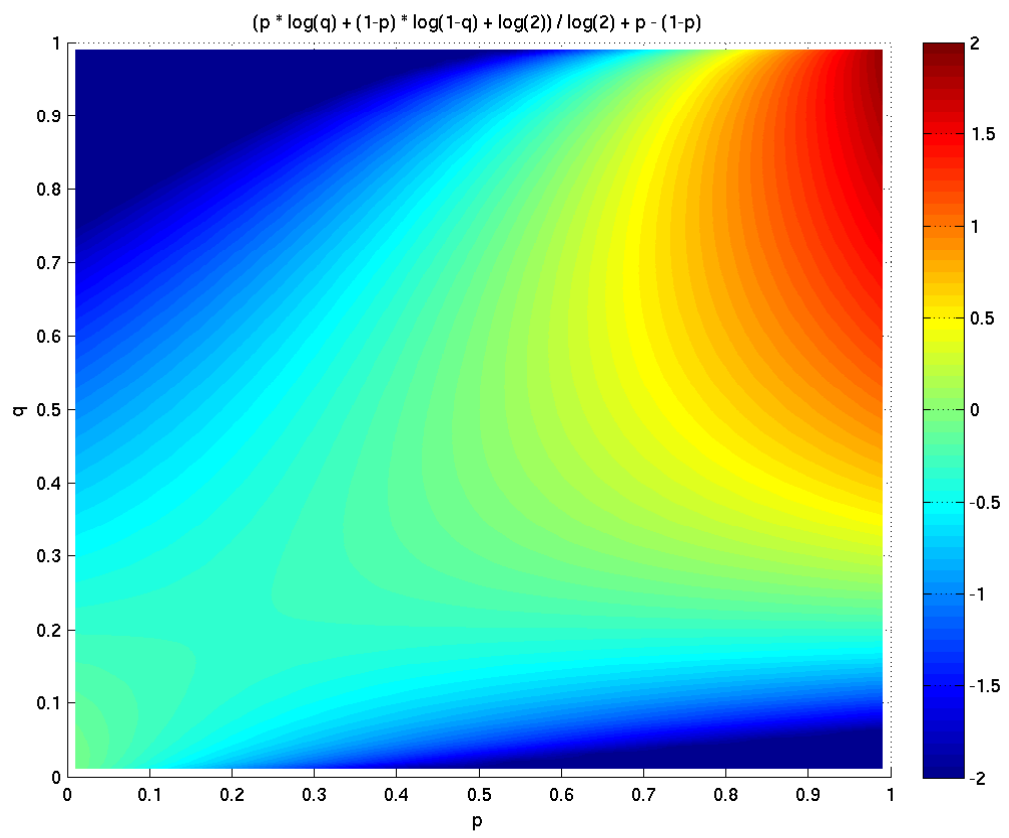}

\caption{Expected total score from scoring +1 for a correct answer, -1
  for a wrong answer, plus the SAC score from figure
  \ref{infoplot}. Note that the colour axis runs from -2 to +2, and
  that the colour scale has been clipped below at -2, with $h$
  approaching $-\infty$ at the top left and bottom right corners.
\label{combined}
}
\end{center}
\end{figure}

One option is to separate out marks for getting the question right and
marks for SAC. If for correctness one scores +1 for a correct answer
and -1 for a wrong answer, while also scoring according to figure
\ref{infoplot} for SAC, the effective combined score resulting is as
in figure \ref{combined}.

This has the disadvantage that if the student thinks the probability
that his answer is correct is less than 0.2, it is slightly in his
interest to change his answer to ensure it is wrong and then say that
he's sure it's wrong.

However, a definite advantage of separating scores for correctness and
SAC in this particular way is that with the method of figure
\ref{infoplot} the student is driven to maximising the apparent
Shannon information content about his own accuracy in his reported
confidence. Moreover, maximising the SAC scores as in figure
\ref{infoplot} also maximises the expected log probability that a
student who marks his own paper randomly, marking correct with the
relevant probability $q$ and wrong with probability $1-q$ on each
question, will mark all the questions correctly. It also provides good
training against overconfidence for the rest of life: if one is
overconfident, one rapidly clocks up some very negative SAC scores.

There are, of course, various psychological factors that could also be
considered when making a choice between the various possible SAC
scoring methods that do maximise score when confidence matches
reality. For example, it can be somewhat discouraging when using the
method of figure \ref{infoplot} to score $-\infty$ for a question -
but it also teaches an important lesson.

Finally we should point out the obvious SAC scoring method for
multiple choice questions with only finitely many choices. In this
instance requiring the student to assign probabilities
$q_1,q_2,...,q_K$ summing to 1 for the various choices, and then
giving him $\log(q_k)$ as his score, where $k$ is the correct answer,
uniquely maximises his expected score when the $q_j$ values match his
subjective probabilities of answer $j$ being correct -- the various
deliberations above are only needed where it is infeasible to assign a
probability distribution over all the possible answers, and instead
concentrate on just right or wrong.

\section{Further analysis of the dataset of Foster's SAC experiment \cite{FosterCA}}

\subsection{Introduction}

In \cite{FosterCA} data was collected from 4 schools. In each school,
students were divided into a control group who were taught by the
traditional method, and an intervention group who were taught
additionally using SAC. Before use of SAC was started, each school ran
a formal test (not using SAC) on all the pupils; these pretests were
the same for each student in a school, but differed between
schools. During and after the period in which SAC was used on one
subset, both subsets underwent further such tests (also not using
SAC); the number of further such tests ranged from 1 to 3 between the
different schools.

The various tests were marked out of various numbers $N_{s,t}$ where
$s$ is the index number of the school and $t$ is the test number in
that school; the values of $N_{s,t}$ ranged from 50 to 131.

Foster both ran a frequentist analysis, which found no significant
difference overall between the performance gain in the control group
and that in the intervention group, and also a Bayesian analysis,
investigating whether one could conclusively say that there was no
difference in the performance gains. He used priors which attempted to
be objective, being based on Jeffreys priors.

However, we would differ from Foster on a few philosophical points, as
follows. 

First, as explained in detail in \cite{WhyBayes}, we believe
frequentist analyses to be frequently misleading, and try to avoid
them and use Bayesian analysis instead.

Second, we think it is usually a mistake to think that a prior can be
``objective''; in particular a Jeffreys prior encapsulates the thought
``I've chosen this experiment because its accuracy profile matches my
prior on the variable being measured''. For example, if I try to
measure current by passing the current in question through a resistive
metal wire and measuring the radiated heat power when it reaches
steady state, and the error in that power measurement is Gaussian with
a standard deviation independent of the radiated power, then the
steady state radiated power is proportional to the square of the
current (assuming resistance stays constant) until such point as the
wire ruptures. Hence the measurement is more sensitive to small
changes at higher current flows than low ones ($\frac{dW}{dI}= 2IR$ if
$W$ is the radiated power, $I$ is the current, and $R$ the
resistance), and the Jeffreys prior will have probability density
proportional to the absolute value of the current up to the rupture
current, after which the Jeffreys prior will be zero. But that might
not be my prior on $I$ at all - it might just be that the only ammeter
I could find was one that worked like this.

A prior can, however, be (at least relatively) uninformative if it
avoids placing near-zero probability density on values of the unknown
that are actually possible.

Third, given two teaching methods, we believe the probability that the
difference in outcome between them will be \textit{exactly} zero will
be zero. For this reason we usually avoid using model comparison that
compares a model in which the difference is always exactly zero with a
model that allows the two to be different; doing so amounts to setting
a prior probability density on some parameter which has an infinite
density spike at the point(s) representing no difference. Instead we
prefer to instead ask what the posterior probability is that the
result of method A is better than that of method B (and assume that
one minus this value represents the probability that method A is worse
than method B). Where we cannot clearly tell whether method A or
method B is better, we note that fact, rather than either assuming or
trying to prove that the two methods give exactly the same results.

Fourth, we believe that Bayes works best when given \textit{all} the
data. Consequently we avoid selecting a subset of the data in one
class that matches that in another class in some way, and instead use
all the data, and look for relationships between the distributions
over the classes and the differences in those distributions when
teaching method is changed.

We therefore set up a rather different probabilistic model from that
in \cite{FosterCA}, and carry out additional analyses to see whether
we can get any further insights on what happened in Foster's very
interesting experiment.

\subsection{The probabilistic model used}

We use a hierarchical Bayesian generative model to describe the system.

By way of notation we define the following:

\begin{itemize}

\item $m$ will denote the teaching method, 1 for traditional and 2 for
  incorporating SAC;

\item $s$ will denote the index number of the school, ranging from 1
  to 4;

\item $t$ will denote the test number within that particular school;
  test 1 is the pretest, and tests 2 to 4 those done after various
  durations of applying the relevant teaching method; 

\item $T_s$ will denote the total number of tests applied in school
  $s$; 

\item $N_{s,t}$ will denote the number of marks allocated in test $t$
  in school $s$;

\item $u$ will denote the index number of the student in his method
  subset of his school;

\item $n_{m,s,t,u}$ will denote the number of marks gained in test $t$
  by student $u$ of method group $m$ in school $s$;

\item $U_{m,s}$ will denote the number of students being taught by
  method $m$ in school $s$;

\item $p_{m,s,t,u}$ will denote the probability that student $u$ of
  the method $m$ subset in school $s$ answers a question in test $t$
  correctly; we assume that this is the same for all questions in that
  test;

\item $K_{m,s,t}$ will denote the number of mixture components in the
  prior on the various $(p_{m,s,t,u})_{u=1,...,U_{m,s}}$;

\item $\lambda$ will denote the parameter of the integer-valued
  exponential prior on each of the $K_{m,s,t}$.

\item $\alpha_{m,s,t,k}, \beta_{m,s,t,k}$ will denote the parameters of the $k$th
  mixture component Beta distribution of the prior on the various
  $(p_{m,s,t,u})_{u=1,...,U_{m,s}}$;

\item $k_{m,s,t,u}$ will be the mixture component number applicable in
  a given sample from the model to student $u$ of the method $m$
  subset in school $s$ for test $t$;

\item $q_{m,s,t,k}$ will be the mixing probability for component $k$;

\item $\gamma_{m,s,t,k}$ will be the corresponding parameter for the
  Dirichlet prior on $(q_{m,s,t,k})_{k=1,...,K_{m,s,t}}$.

\item $\mu$ is a parameter of the prior on the vector
  $(\gamma_{m,s,t,k})_{k=1,...,K}$ given $K_{m,s,t}$.

\item $\kappa,a,b$ are the parameters of the proBeta prior on the various
  $\alpha_{m,s,t,k},\beta_{m,s,t,k}$.

\end{itemize}

We assume that a student's marks in a test are binomially distributed
with parameters $N_{s,t},p_{m,s,t,u}$. We put a prior on each element
of the set $\{p_{m,s,t,u}:u\in\{1,...,U_{m,s}\}\}$ that is a mixture of
$K_{m,s,t}$ Beta distributions with mixing probabilities
$q_{m,s,t,k}\geq 0$ summing to 1 over $k=1,...,K_{m,s,t}$. The
parameters of these Beta distributions being $\alpha_{m,s,t,k},
\beta_{m,s,t,k}$, we put a proBeta prior\footnote{The conjugate
distribution to the joint parameters $(\alpha,\beta)$ of the Beta
distribution.} on those given for all values of $(m,s,t,k)$
independently by $$P(\alpha,\beta|\kappa,a,b) \propto
\left(\frac{\Gamma(\alpha+\beta)}{\Gamma(\alpha)\Gamma(\beta)}\right)^\kappa
a^{\kappa(\alpha-1)}b^{\kappa(\beta-1)},$$ for $\alpha,\beta>0$ where
$\kappa>0,a>0,b>0,a+b<1$. We put a Dirichlet prior on the vector
$(q_{m,s,t,k})_{k=1,...,K_{m,s,t}}$ with parameter vector
$(\gamma_{m,s,t,k})_{k=1,...,K_{m,s,t}}$, given for each value of
$(m,s,t)$ independently by $$P(q|\gamma) =
\frac{1}{\sqrt{K}}\frac{\Gamma(\sum_{k=1}^K{\gamma_k})}
     {\prod_{k=1}^K{\Gamma(\gamma_k)}}\prod_{k=1}^K{q_k^{\gamma_k-1}}$$
     for $\sum_{k=1}^K{q_k}=1$ and all $q_k>0$. We put a prior on the
     vector $(\gamma_{m,s,t,k})_{k=1,...,K}$ given $K_{m,s,t}$ putting all the
     probability on the single value $\mu/K_{m,s,t}$. 
     Finally we put an integer-valued exponential prior on the number
     of mixture components, i.e. on each $K_{m,s,t}$ independently,
     given by $$P(K|\lambda) = \frac{\lambda^{-(K-1)}}{1-\lambda}$$
     for $\lambda\in (0,1)$ and $K = 1,2,...$ . 

Thus our probabilistic model may be defined by the following
hierarchical equations:

$$\kappa,a,b,\lambda,\mu\text{ are constants defining the prior}$$
$$\forall (m,s,t),\ K=K_{m,s,t}:
P(K|\lambda)=\frac{\lambda^{-(K-1)}}{1-\lambda}\ \ (K=1,2,3,...)$$
$$\forall (m,s,t,k),\ (\alpha,\beta)=(\alpha_{m,s,t,k},\beta_{m,s,t,k}): P(\alpha,\beta|\kappa,a,b) \propto
\left(\frac{\Gamma(\alpha+\beta)}{\Gamma(\alpha)\Gamma(\beta)}\right)^\kappa
a^{\kappa(\alpha-1)}b^{\kappa(\beta-1)}\ \ (\alpha,\beta>0)$$
$$\forall (m,s,t,k), \gamma_{m,s,t,k}=\mu/K_{m,s,t}$$
$$\forall (m,s,t), q=(q_{m,s,t,k})_{k=1,...,K_{m,s,t}},\gamma=(\gamma_{m,s,t,k})_{k=1,...,K_{m,s,t}}:\ P(q|\gamma) =
  \frac{1}{\sqrt{K}}\frac{\Gamma(\sum_{k=1}^K{\gamma_k})}
       {\prod_{k=1}^K{\Gamma(\gamma_k)}}\prod_{k=1}^K{q_k^{\gamma_k-1}}\ \ (q_k>0,\sum_{k=1}^K{q_k}=1)$$
$$\forall (m,s,t,u),
       k=k_{m,s,t,u},q=(q_{m,s,t,k})_{k=1,...,K_{m,s,t}}:\ P(k|q)=q_{k}\ \ (k=1,...,K_{m,s,t})$$
$$\forall (m,s,t,u), p =
       p_{m,s,t,u},\alpha=\alpha_{m,s,t,k_{m,s,t,u}},\beta=\beta_{m,s,t,k_{m,s,t,u}}:\ P(p|\alpha,\beta)
       = \frac{\Gamma(\alpha+\beta)}{\Gamma(\alpha)\Gamma(\beta)}
       p^{\alpha-1}(1-p)^{\beta-1}\ \ (0<p<1)$$
$$\forall (m,s,t,u),
       n=n_{m,s,t,u},N=N_{m,s,t},p=p_{m,s,t,u}:\ P(n|N,p) =
       \frac{N!}{n!(N-n)!}p^n(1-p)^{N-n}\ \ (n=0,1,...,N).$$

\subsection{Priors used}

\begin{figure}[hpt]
\begin{center}
\includegraphics[scale=0.5]{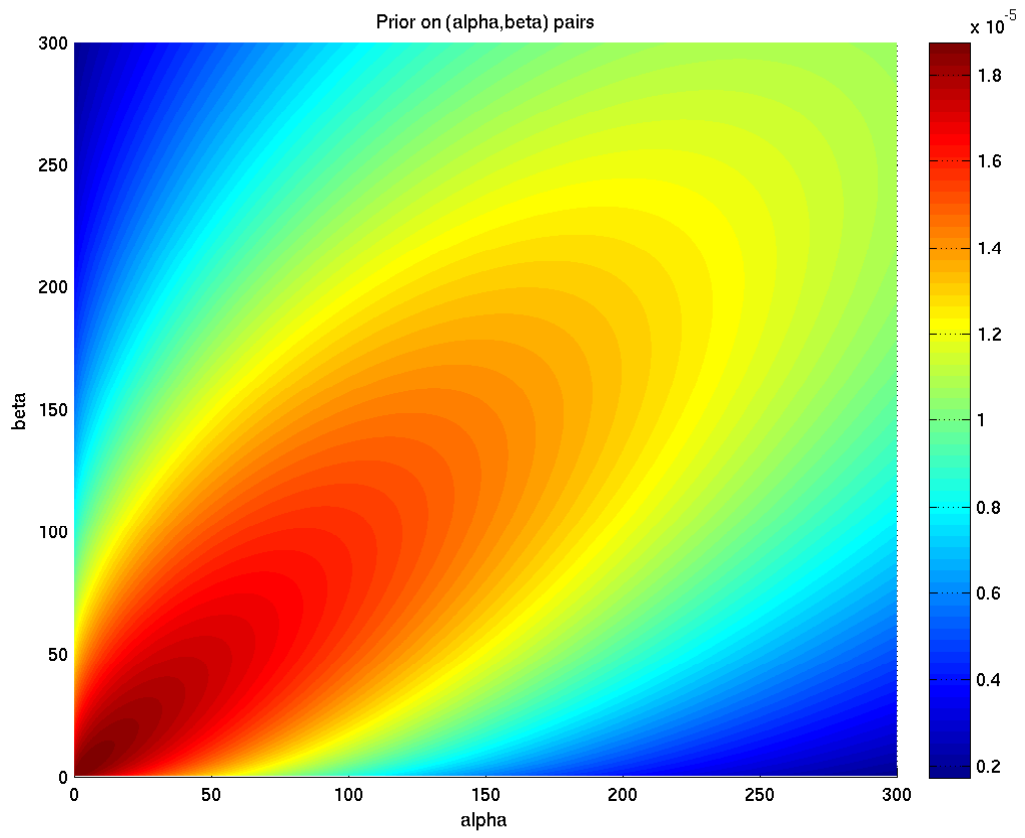}

\caption{proBeta prior used for each
  $(\alpha_{m,s,t,k},\beta_{m,s,t,k})$ pair.
\label{proBeta}
}
\end{center}
\end{figure}

We set the following values of the top level parameters: $$\kappa =
0.01; a=b=0.4525; \lambda = 0.5; \mu = 1.$$ Thus we put most of the
density of the Dirichlet prior on the mixture coefficients near the
edges of the $K$-simplex (any individual mixture component is likely
to have small weight), and the proBeta prior on the $(\alpha,\beta)$
pairs has the density shown in figure \ref{proBeta}.

\subsection{Exploration of the model}

We explore the posterior distribution of the model given the observed
data $U,T,N,n$ and the priors $\kappa,a,b,\lambda,\mu$ (i.e. we take
joint samples from the values of all the other parameters) using
Markov chain Monte-Carlo sampling (see \cite{rNealMCMC}). Specific
details are as follows:

\begin{itemize}

\item We use Gibbs sampling, visiting the variables palindromically
  moving from bottom to top of the model and back again, repeatedly.

\item We integrate out all the variables $p_{m,s,t,u}$ to increase
  mobility, then at the end resample them all given their parents and
  children in the model to complete our sample set.

\item The proBeta distribution is log-concave, so we use adaptive
  rejection sampling\cite{GilksARMS} along orthogonal
  directions in $(\alpha,\beta)$ space, one direction being on a
  straight line through the origin and our current point.

\item For the methods needed to sample the remaining conditional
  distributions that occur, see \cite{Dagpunar}.

\end{itemize}

Having drawn a suitably large number (e.g. 10,000) samples of the
vector of all variables in the system, we can now ask detailed
questions of the model. For example, we can easily produce, for each
sample and each student, the increase in log-odds of getting a
question right from pretest to one of the posttests, or for each
sample the average such gain in each (school, posttest, method group)
subset. These samples then represent the posterior distribution of
each such quantity, and allow us to calculate, for each school and
posttest, the probability that that gain is bigger for one method than
for the other, and the average value of that gain.

Alternatively, we can repeat the above analysis but restricting
attention to the upper half or top quartile of each class, or to the
lower half or bottom quartile, to see whether the change in method
works better with strong or weak students.

Alternatively, we can restrict attention e.g. to a particular school
and posttest combination, and plot for each student the probability
that their posttest log-odds are higher than their pretest log-odds
against their position in class at pretest.

Numerous other inferences can similarly be drawn, and we illustrate
some of them in section \ref{Results} below.

\subsection{Software testing and induced priors}

\begin{figure}
\begin{tabular}{cc}

\includegraphics[scale=0.5]{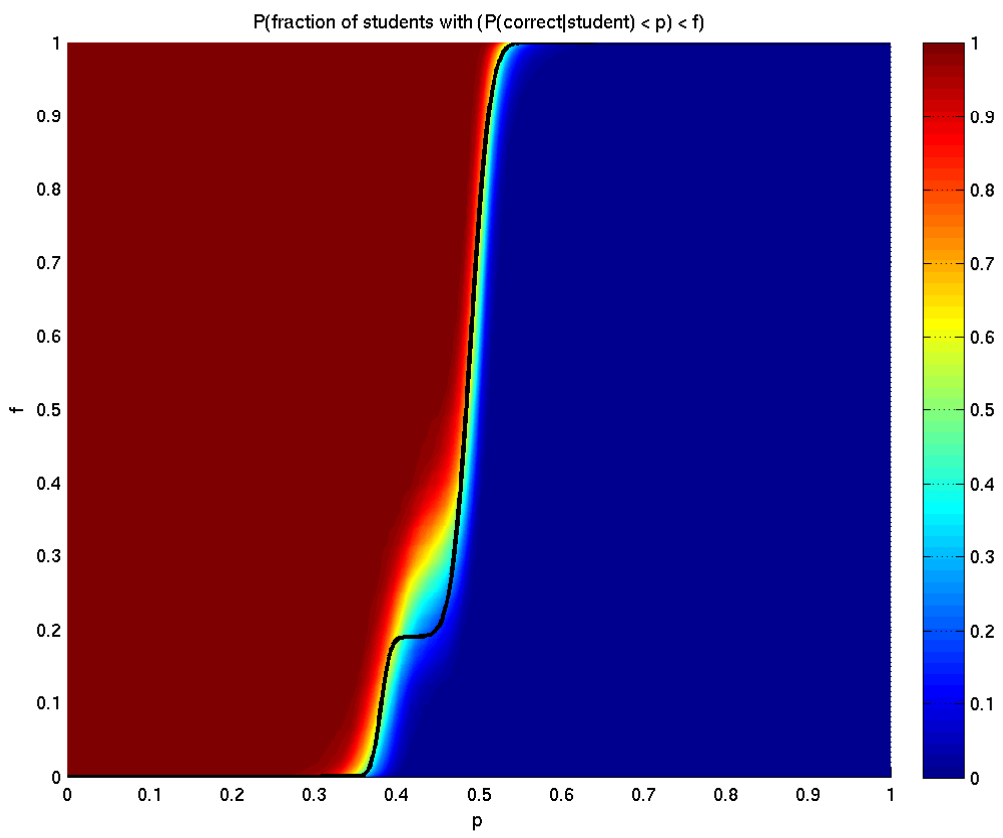} &
\includegraphics[scale=0.5]{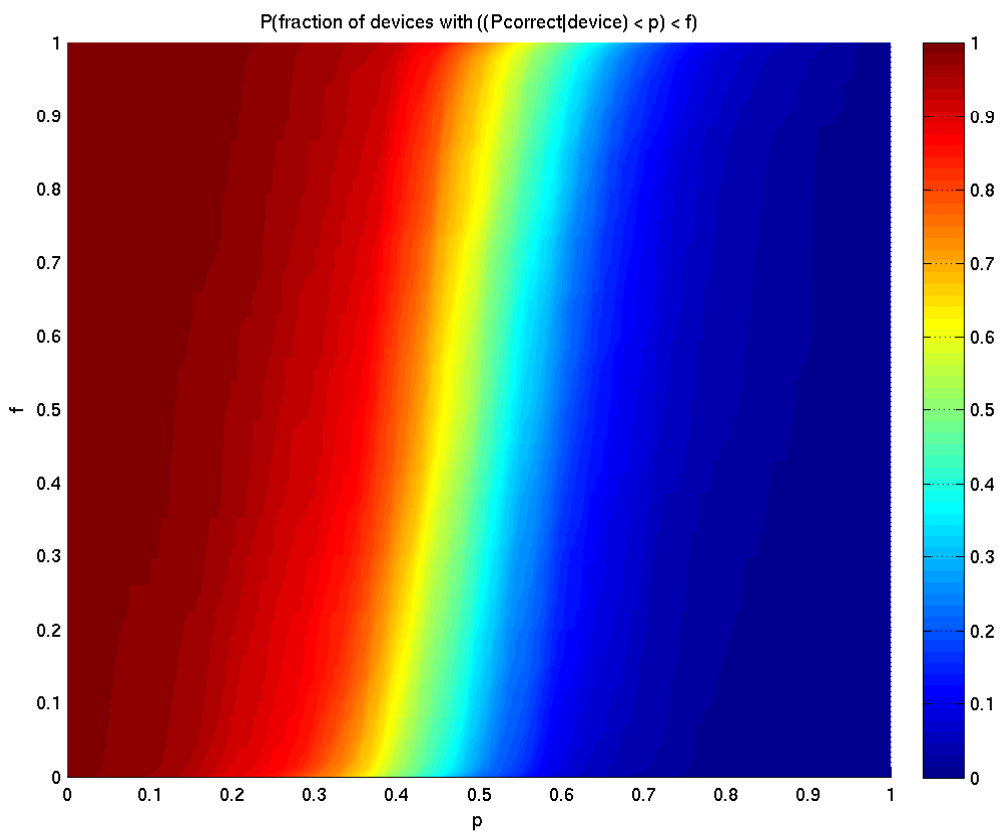} \\
\includegraphics[scale=0.5]{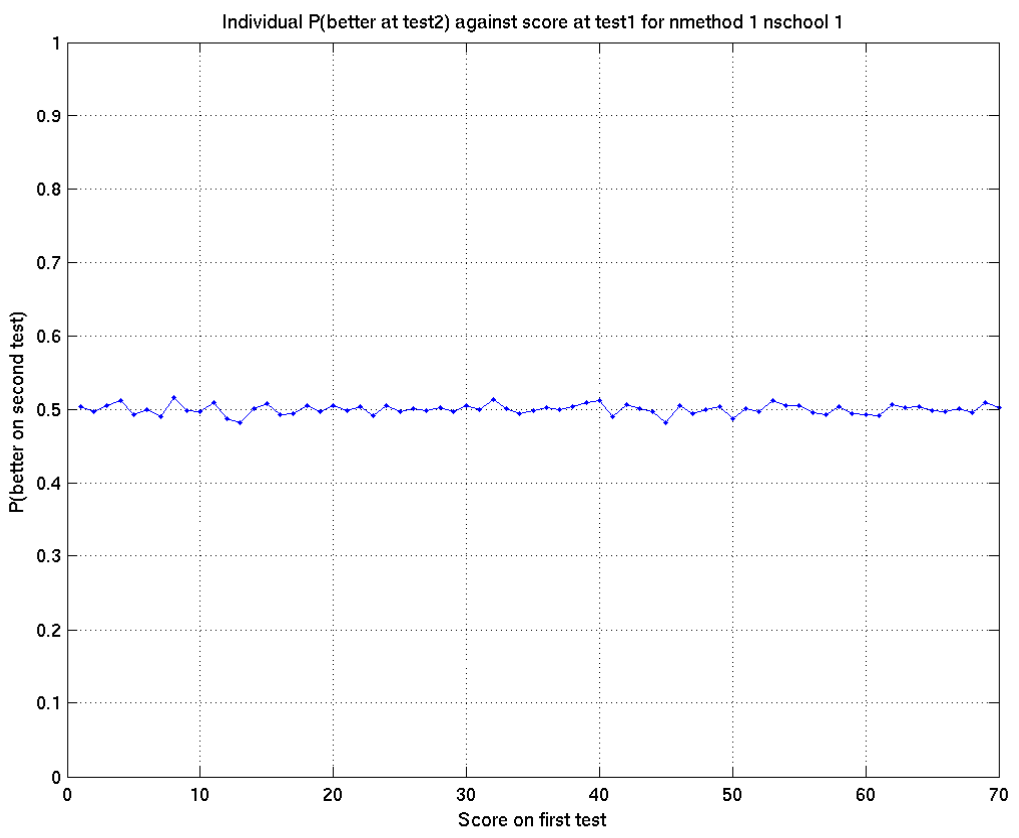} &
\includegraphics[scale=0.5]{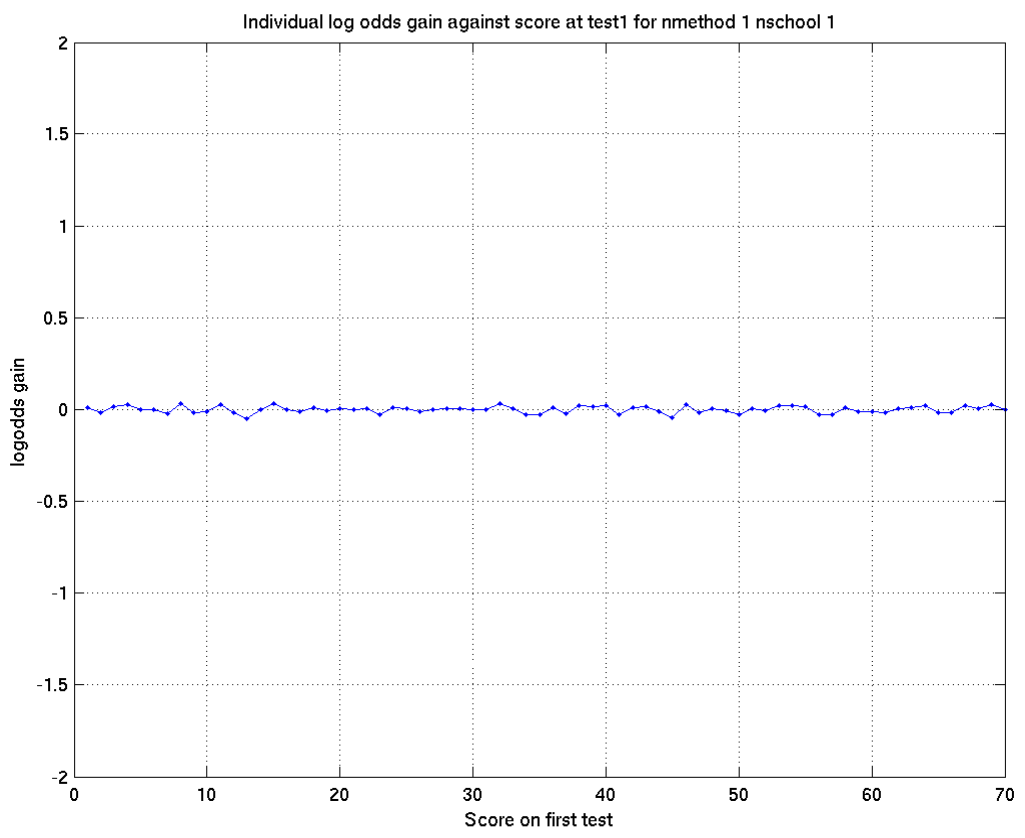} \\

\end{tabular}

\caption{Plots confirming correct operation of the software and
  sensible settings for the priors after running the Markov chain
  Monte-Carlo system for 10,000 samples. The top left plot uses
  synthetic data, for which the truth is known; the true cumulative
  distribution function (cdf) of the probability of a student in this
  hypothetical class getting a question correct is shown as the black
  line, while the posterior distribution of this cdf is shown as the
  background colour plot. The top right plot shows similar inference
  made using a test consisting of zero questions, so that it reflects
  the induced prior on this cdf. The bottom left plot shows that when
  two such tests consisting of zero questions are applied to a class,
  the probability of the inter-test gain in log-odds of individual
  students getting a question correct being positive is almost exactly
  0.5 for every student, and the bottom right plot shows that the
  posterior expectation of gain in log-odds is very close to zero. In
  both the bottom plots the ``score on first test'' is just set to the
  index-number of the student, as obviously all score zero on a test
  consisting of zero questions.
\label{testplot}
}
\end{figure}

To confirm correct operation of the software we ran it on synthetic
data, getting the inference shown in the top left plot of figure
\ref{testplot} (of the same type as those shown in figures
\ref{school2cdf} and \ref{school4cdf}); in this case the truth is
known and is also shown, lying comfortably within the inferred
distribution. To visualise the induced prior on this inference, we ran
the software using two classes of 70 students each doing a pretest and
a posttest each consisting of zero questions; this gave the top right
plot of figure \ref{testplot} for pretest on the ``non-SAC'' class,
showing that the induced prior on the distribution over the class is
wide; similar plots (not shown) were obtained for the other class and
on both posttests. Under the prior the probability of an increase in
gain of log-odds of getting a question correct due to SAC was 0.495
(using only 10,000 samples the difference from 0.5 is not surprising),
and the expected increase in gain was 0.004 nats (for comparison with
tables \ref{wholeclass}, \ref{halves}, and \ref{quartiles}). The
bottom left plot of figure \ref{testplot} shows that under the prior
the probability of log-odds of getting a question correct increasing
from pretest to posttest is almost exactly 0.5 for every student (for
comparison with figure \ref{indivprobs}), and typical gains for
individual students are very small as shown in the bottom right plot
(for comparison with figure \ref{indivgains}).

Thus the software appears to be correctly working and the prior is
inducing sensible distributions on dependent variables.

\subsection{Results}
\label{Results}

For simplicity we only report results comparing the pretest with the
final posttest. We will use the word ``class'' to mean one of the
subsets of pupils in a single school who were taught by the same
method (SAC or no SAC), even though some of these consisted of several
classes as commonly understood.

\subsubsection{Differences in mean increase over class in log-odds of answer
  correct}
\label{meanoverclassparts}

We first report on the class as a whole, i.e. taking all those at a
school given SAC compared with all those not given SAC. Table
\ref{wholeclass} shows that none of these results reaches the 0.05 or
0.95 probability level, and that the increases in gain from pretest to
postest in log-odds of getting a correct answer to a question are all
less than 0.1 nats in magnitude. These results are not surprising
given the findings in \cite{FosterCA}.

\begin{table}[hpt]
\begin{center}
\begin{tabular}{c|c|c}
School & $P$(method 2 gain $>$ method 1 gain) & E(method 2 gain $-$ method
1 gain)\\ & & (nats)\\ \hline
1 & 0.596 & +0.027\\
2 & 0.725 & +0.041\\
3 & 0.145 & -0.082\\
4 & 0.238 & -0.081\\ \hline
\end{tabular}
\caption{For whole class, posterior probabilities that the gain in
  log-odds of answer correct from pretest to posttest is better with
  SAC than without, and the expected gain increase by using SAC.}
\label{wholeclass}
\end{center}
\end{table}

However, just saying this doesn't tell us all that this data
contains. If we instead split each class into two, with those scoring
most in pretest in the ``a'' part and those scoring least in the ``b''
part, we get table \ref{halves}. Now we see that there are quite large
effects emerging, but that these differ in the various schools: in
school 2 introduction of SAC favours the top half of the class, while
in school 4 it favours the bottom half of the class. If we divide into
quartiles we get table \ref{quartiles}, where we see similar trends.

\begin{table}[hpt]
\begin{center}
\begin{tabularew}{c*{4}{|>{\spew{.25}{+1}}c<{\spew{.25}{+1}}}}
School & \multicolumn{2}{c|}{$P$(method 2 gain $>$ method 1 gain)} &
  \multicolumn{2}{c}{E(method 2 gain $-$ method 1 gain) (nats)}\\
& a & b & a & b\\ \hline
1 & 0.796 & 0.270 & +0.080 & -0.058 \\
2 & 0.979 & 0.081 & +0.226 & -0.252 \\
3 & 0.392 & 0.091 & -0.031 & -0.138 \\
4 & 0.005 & 0.946 & -0.400 & +0.202 \\ \hline
\end{tabularew}
\caption{For class divided into two by pretest score (a high, b low),
  posterior probabilities that the gain in log-odds of answer correct from
  pretest to posttest is better with SAC than without, and the
  expected gain increase by using SAC.}
\label{halves}
\end{center}
\end{table}

\begin{table}[hpt]
\begin{center}
\begin{tabularew}{c*{8}{|>{\spew{.25}{+1}}c<{\spew{.25}{+1}}}}
School & \multicolumn{4}{c|}{$P$(method 2 gain $>$ method 1 gain)} &
  \multicolumn{4}{c}{E(method 2 gain $-$ method 1 gain) (nats)}\\
& a & b & c & d & a & b & c & d \\ \hline
1 & 0.699 & 0.586 & 0.234 & 0.260 & +0.077 & -0.005 & -0.087 & -0.082 \\
2 & 0.854 & 0.999 & 0.775 & 0.132 & +0.172 & +0.370 & +0.075 & -0.122 \\
3 & 0.567 & 0.200 & 0.406 & 0.011 & +0.027 & -0.145 & -0.033 & -0.329 \\
4 & 0.000 & 0.295 & 0.941 & 0.962 & -0.661 & -0.099 & +0.246 & +0.285 \\ \hline
\end{tabularew}
\caption{For class divided into four by pretest score (a high,..., d
  low), posterior probabilities that the gain in log-odds of answer correct from
  pretest to posttest is better with SAC than without, and the
  expected gain increase by using SAC.}
\label{quartiles}
\end{center}
\end{table}

\subsubsection{Individual students' gains in log-odds of answer
  correct}

We can also look at what happens with individual students in these
classes, and how how they fared varied with their rank in the pretest.

\begin{figure}
\begin{tabular}{cc}

\includegraphics[scale=0.5]{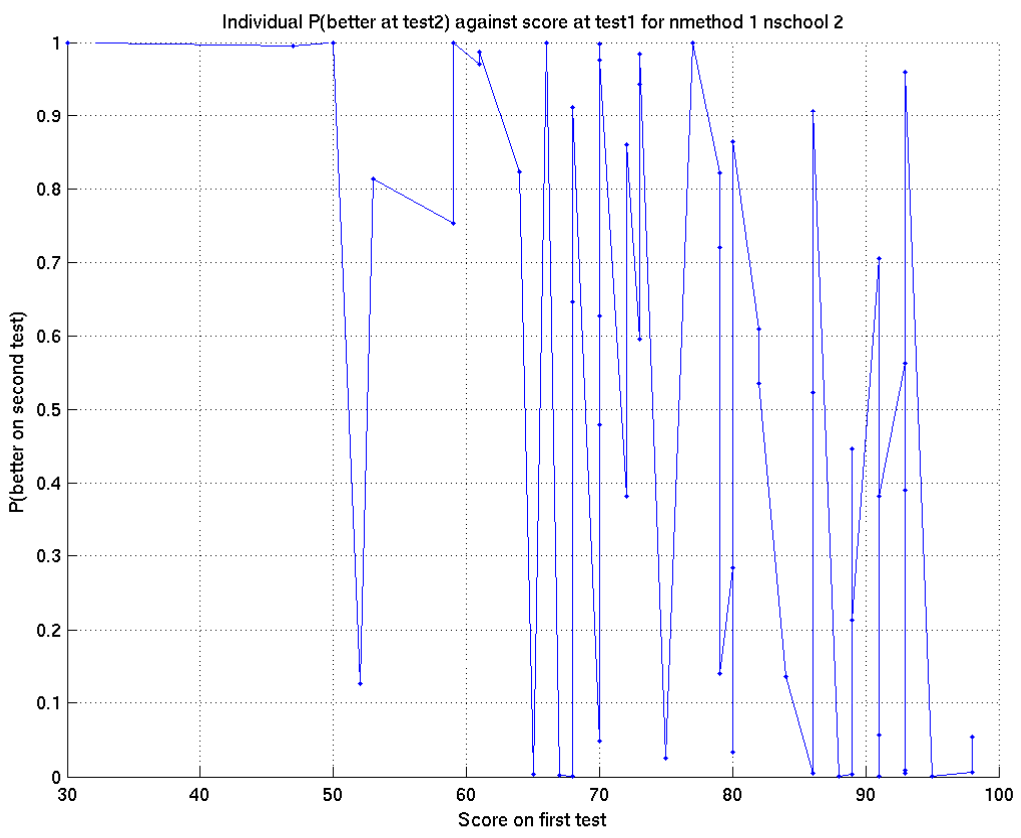} &
\includegraphics[scale=0.5]{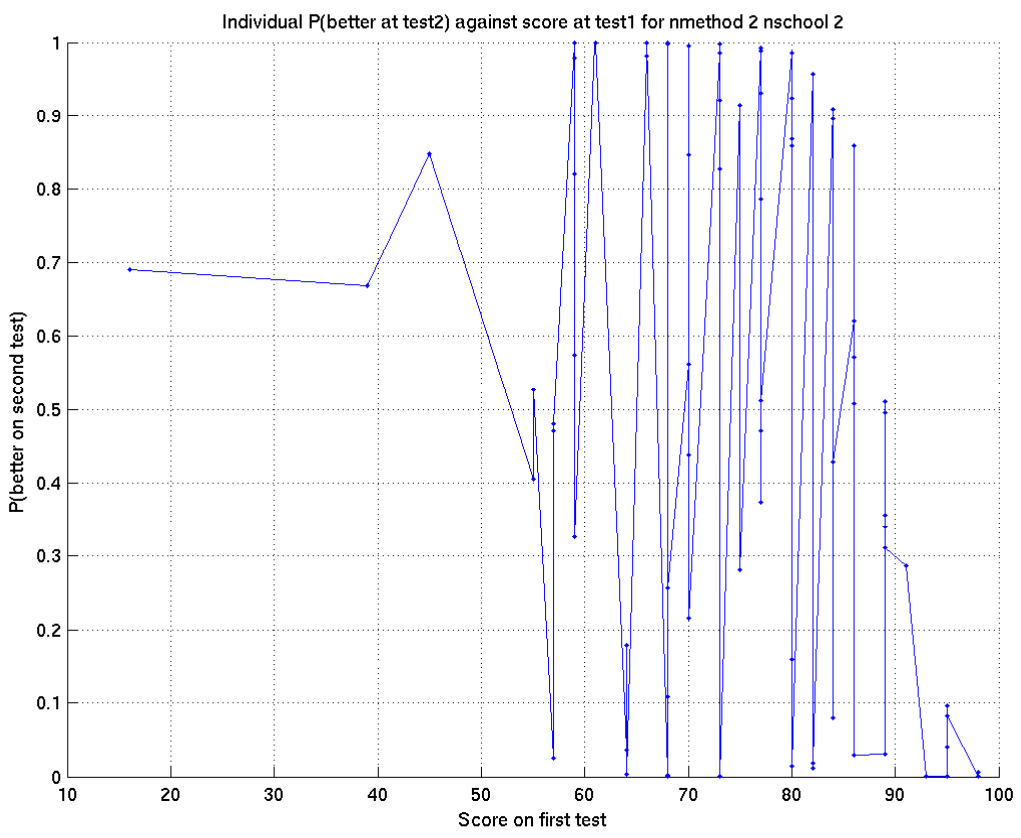} \\
\includegraphics[scale=0.5]{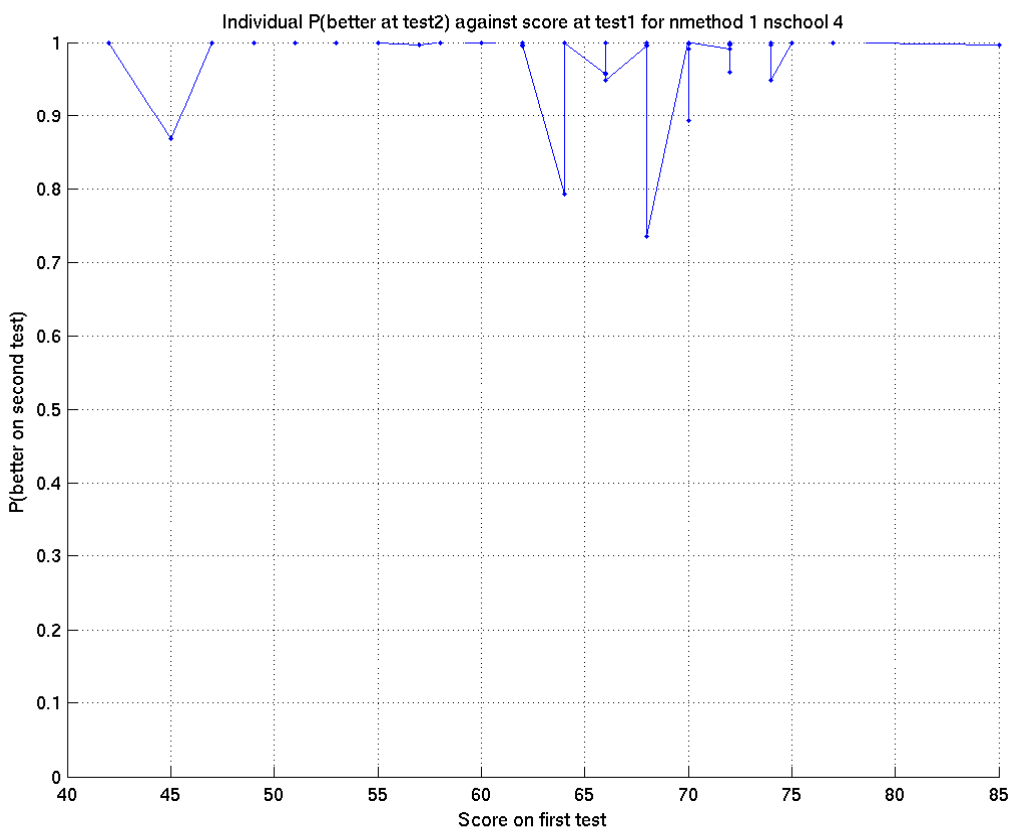} &
\includegraphics[scale=0.5]{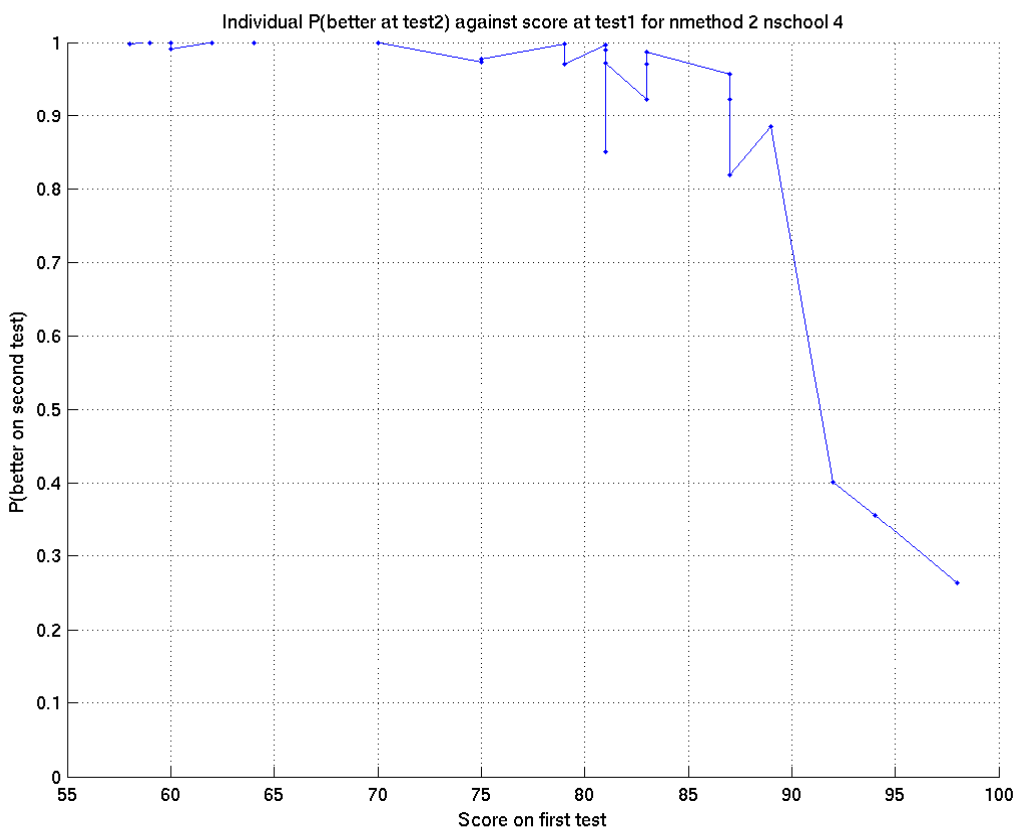} \\

\end{tabular}

\caption{Posterior probabilities for individual students that their
  probability of answering correctly increased from pretest to
  posttest plotted against their mark in the pretest. The top two
  plots are for school 2, in whom SAC appeared to reduce gain in the
  bottom quartile, and the bottom two plots are for school 4, in whom
  SAC appeared to increase gain in the bottom quartile. The left hand
  plots are without SAC and the right hand plots are with SAC Note
  that the set of students is different in each plot, and that the
  tests done differ in the top row and bottom row.
\label{indivprobs}
}
\end{figure}

\begin{figure}
\begin{tabular}{cc}

\includegraphics[scale=0.5]{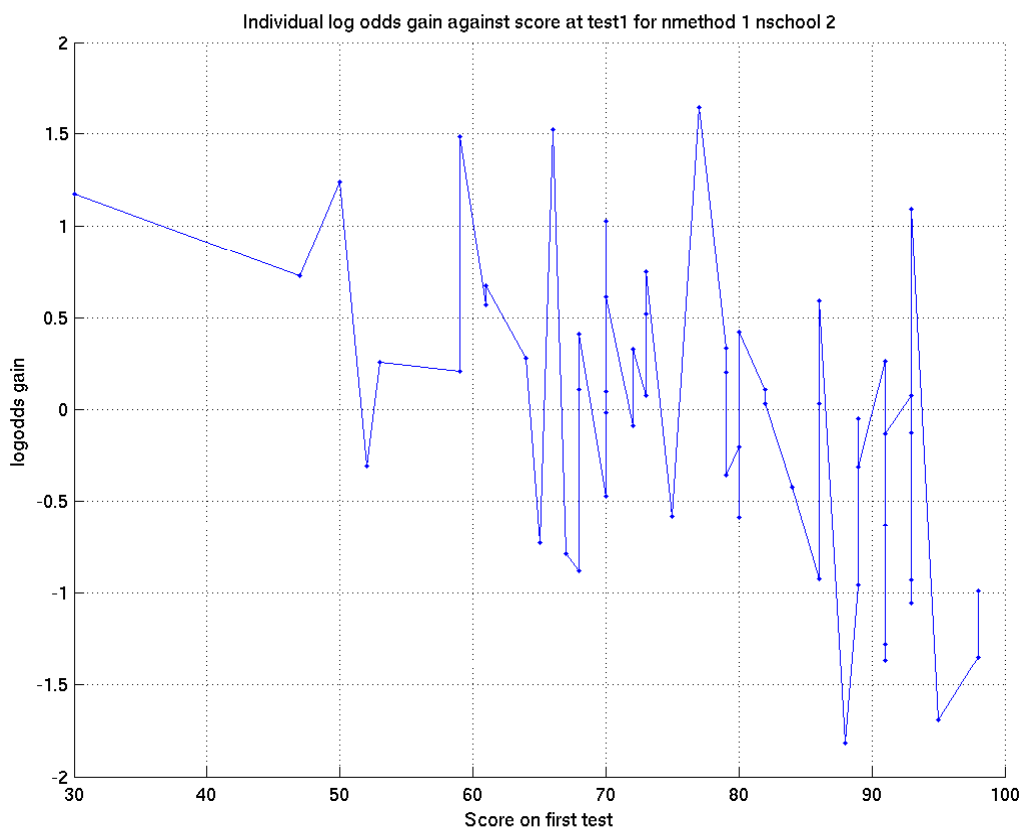} &
\includegraphics[scale=0.5]{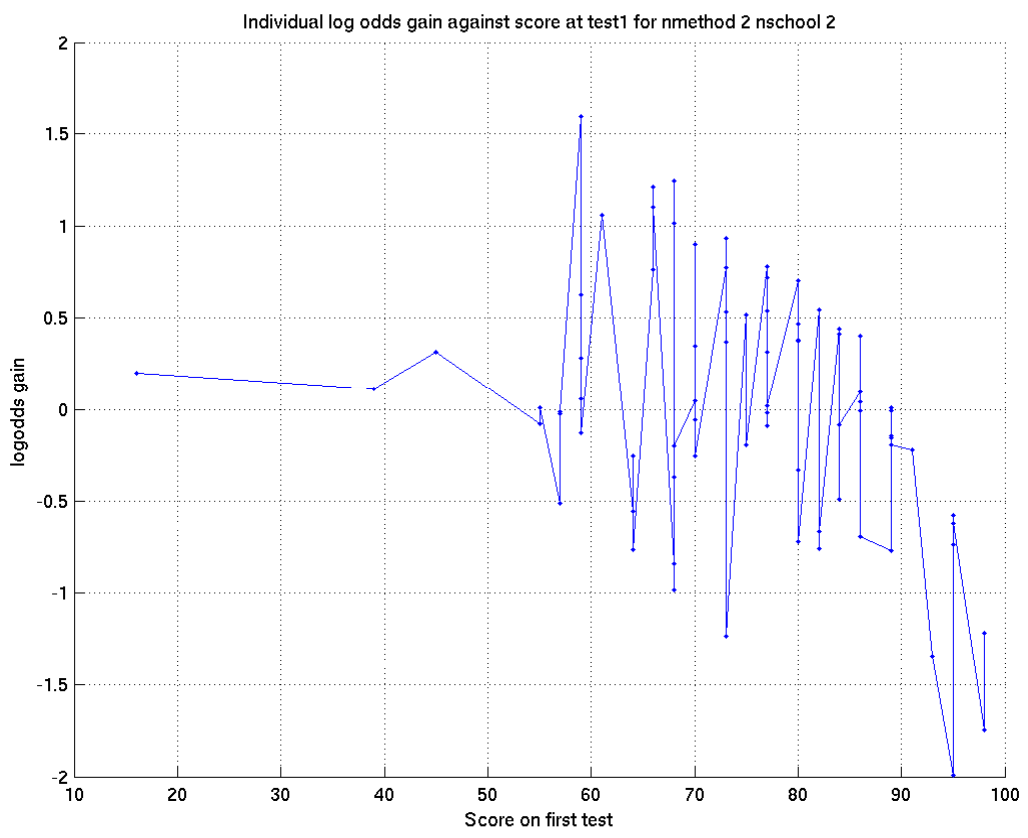} \\
\includegraphics[scale=0.5]{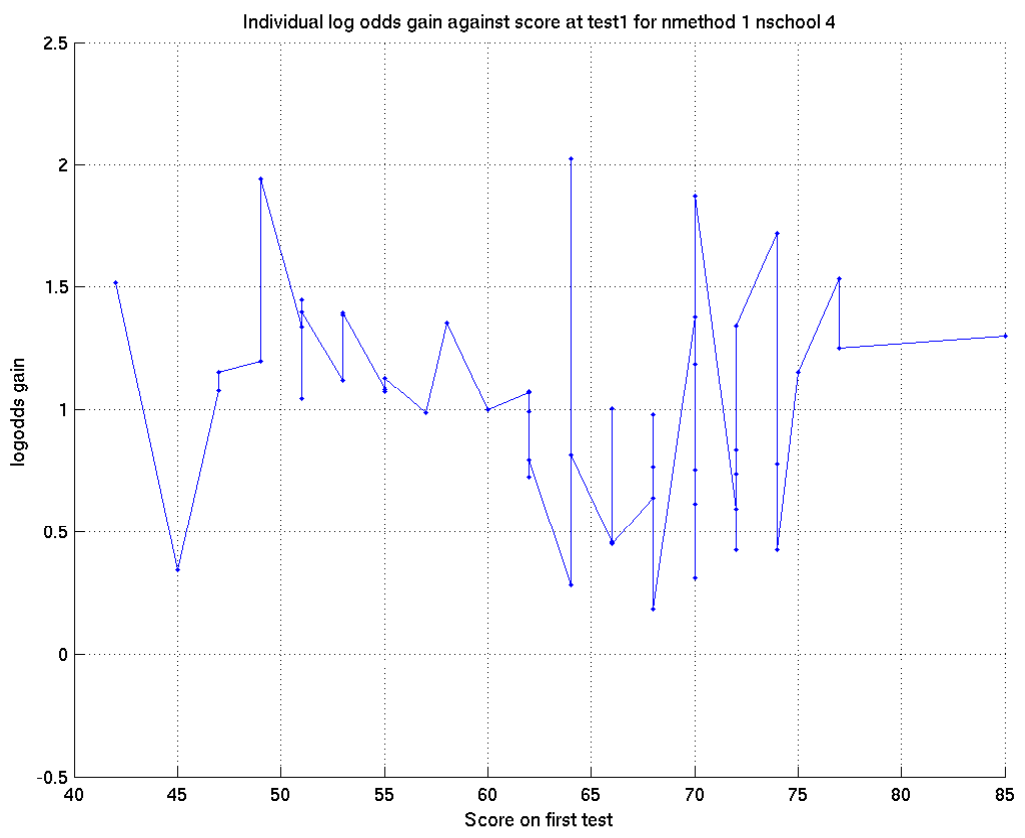} &
\includegraphics[scale=0.5]{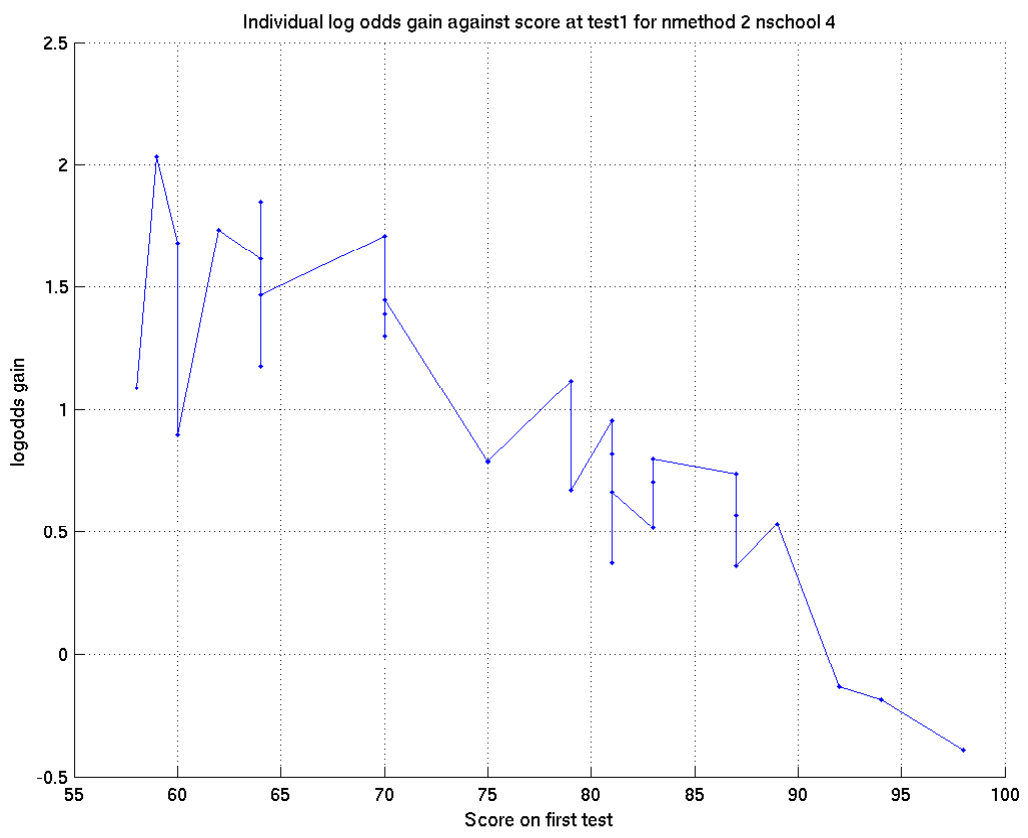} \\

\end{tabular}

\caption{Posterior expected gain in log-odds of answering correctly
  from pretest to posttest for individual students plotted against
  their mark in the pretest. The top two plots are for school 2, in
  whom SAC appeared to reduce gain in the bottom quartile, and the
  bottom two plots are for school 4, in whom SAC appeared to increase
  gain in the bottom quartile. The left hand plots are without SAC and
  the right hand plots are with SAC; Note that the set of students is
  different in each plot, and that the tests done differ in the top
  row and bottom row.
\label{indivgains}
}
\end{figure}

These plots (figures \ref{indivprobs} and \ref{indivgains}) show a
degree of consistency with the results in section
\ref{meanoverclassparts}, in that in school 2 introduction of SAC
appears to have been bad for those at the bottom the class, while in
school 4 introduction of SAC appears to have been bad for those at the
top of the class and good for those at the bottom. We note that this
is not \textit{just} an observation made after averaging out across
students, but appear to be consistent within classes for those at the
top and bottom.

\subsubsection{Pretest to posttest changes in distribution over class
  of log-odds of answer correct}

\begin{figure}
\begin{tabular}{cc}

\includegraphics[scale=0.5]{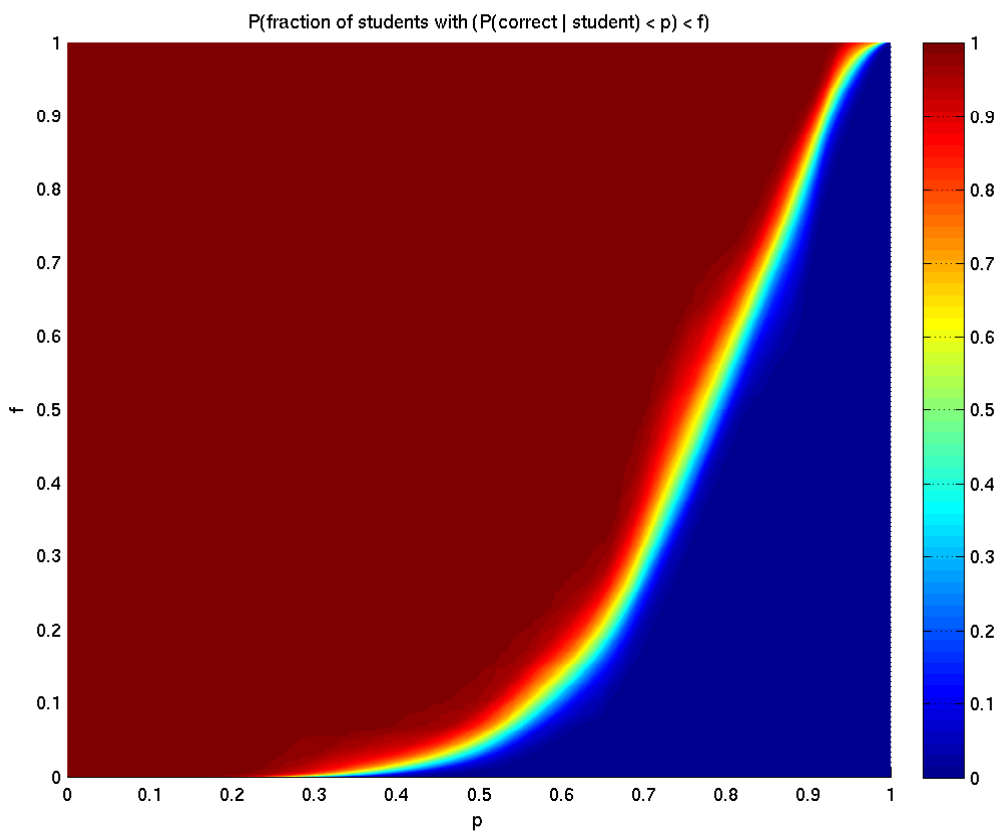} &
\includegraphics[scale=0.5]{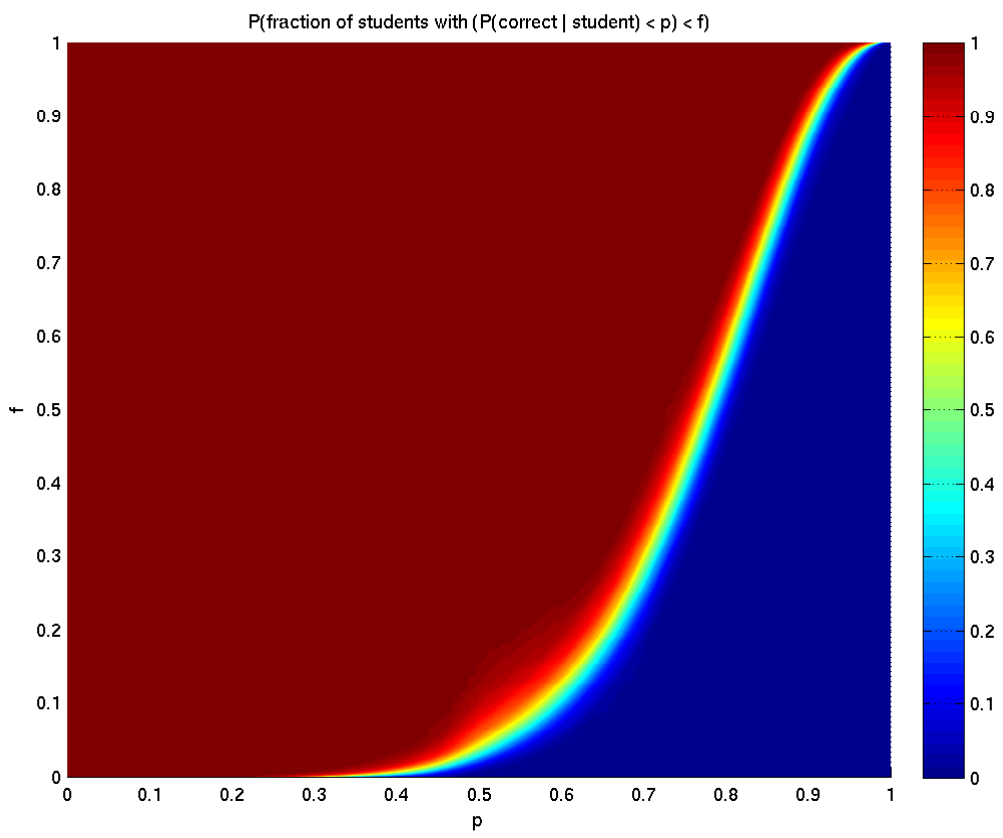} \\
\includegraphics[scale=0.5]{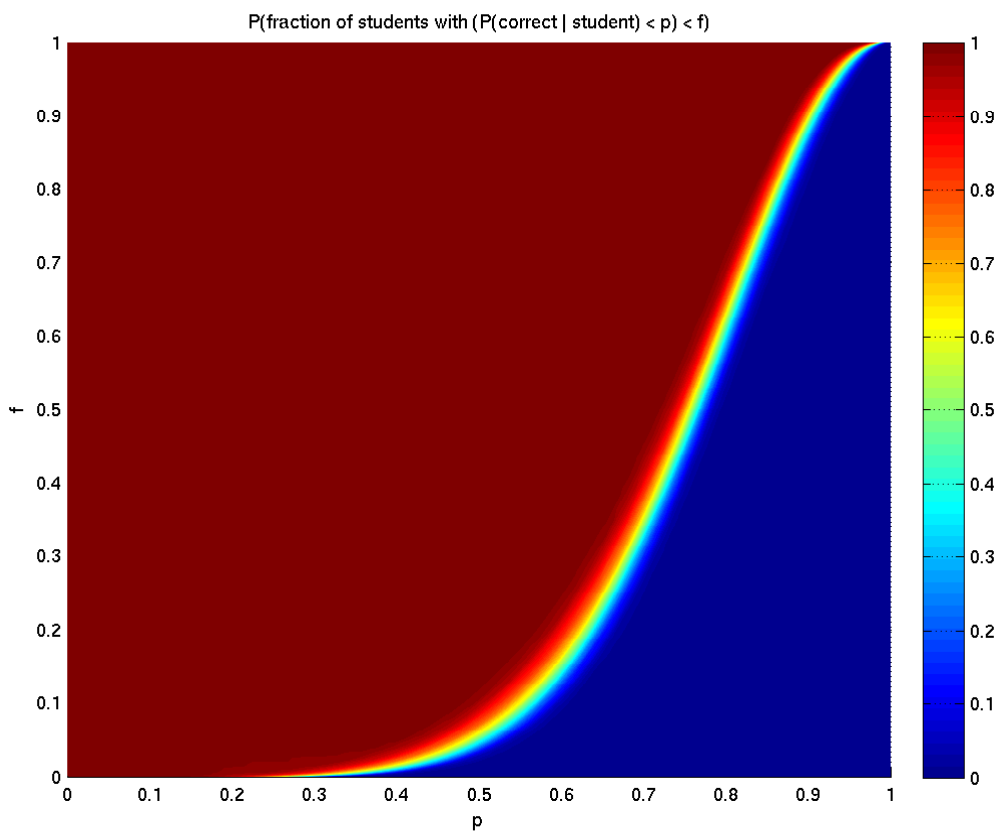} &
\includegraphics[scale=0.5]{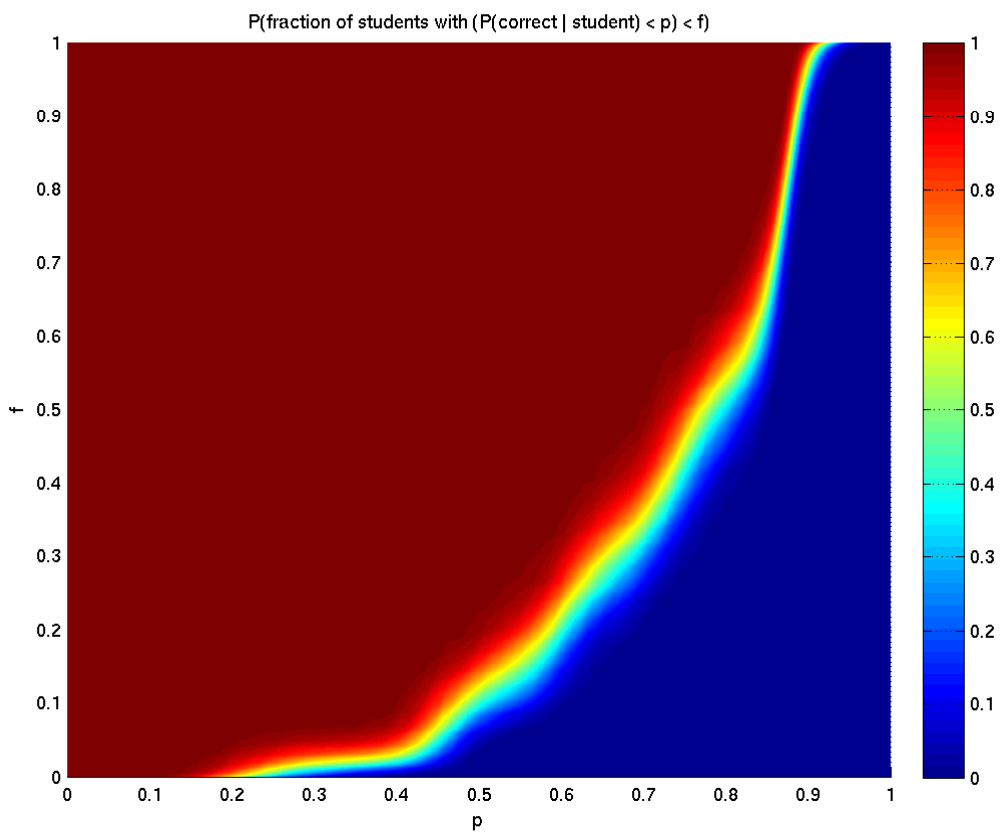} \\

\end{tabular}

\caption{Posterior distribution of the cumulative distribution
  function over students in each class of the probability of getting a
  question correct: school2. The left two plots are for the pretest
  and the right two plots for the posttest. The top two plots are
  without SAC and the bottom two with SAC.
\label{school2cdf}
}
\end{figure}

\begin{figure}
\begin{tabular}{cc}

\includegraphics[scale=0.5]{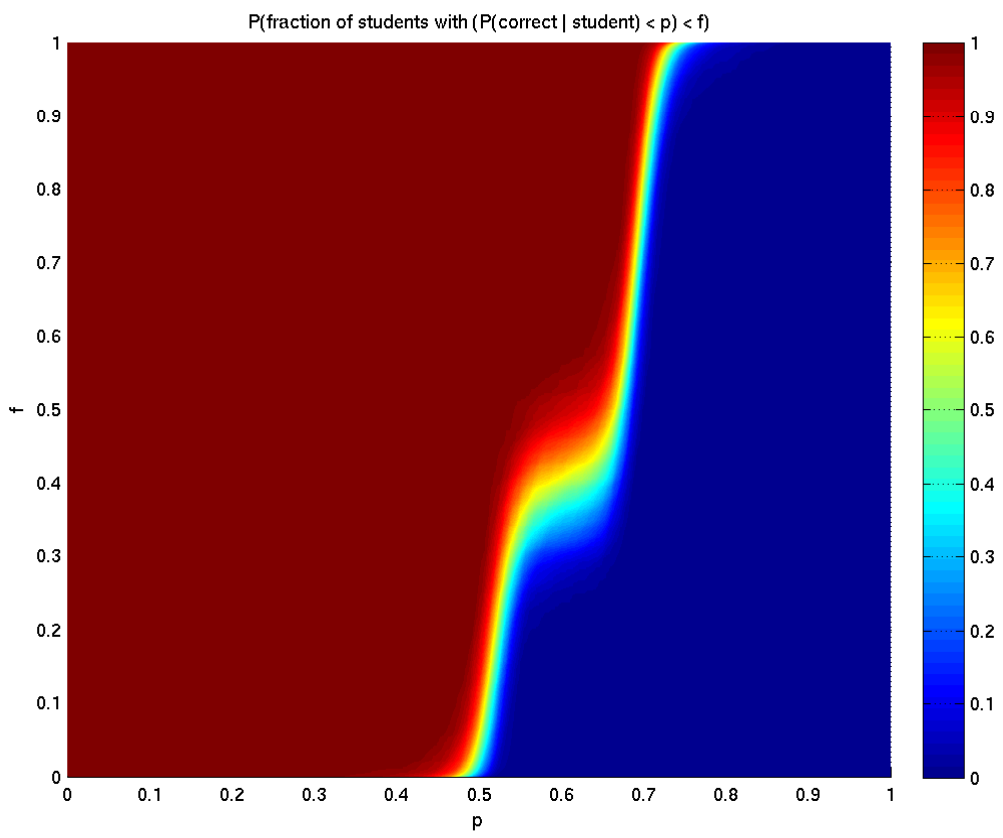} &
\includegraphics[scale=0.5]{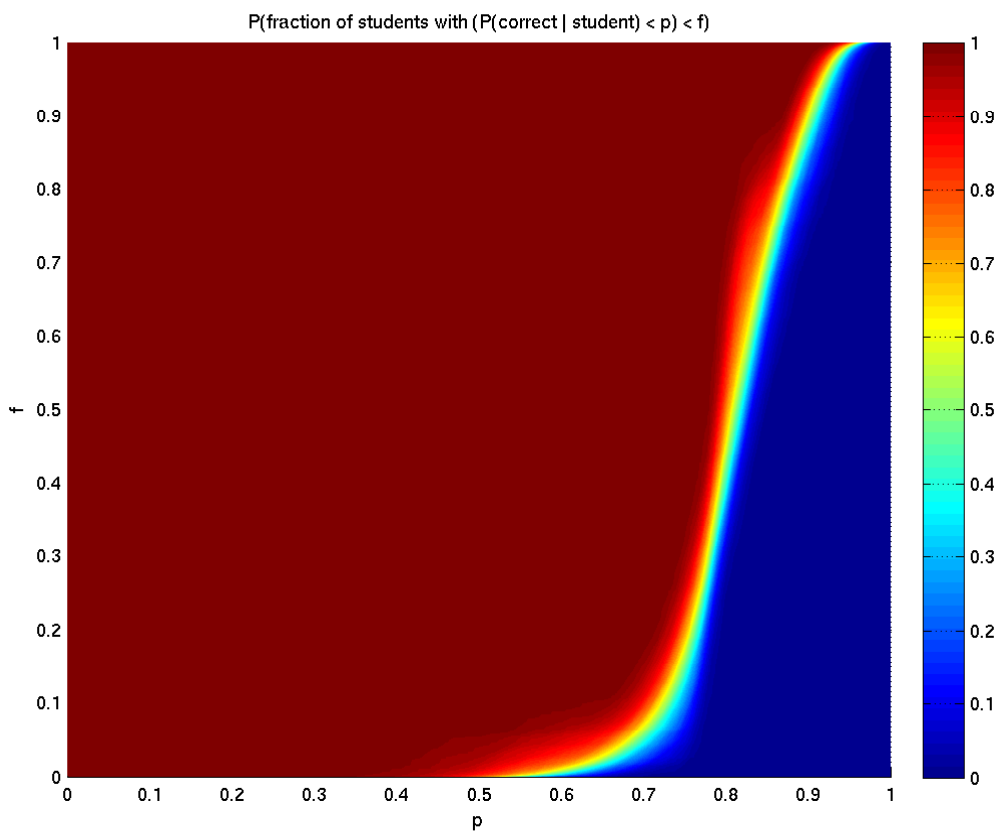} \\
\includegraphics[scale=0.5]{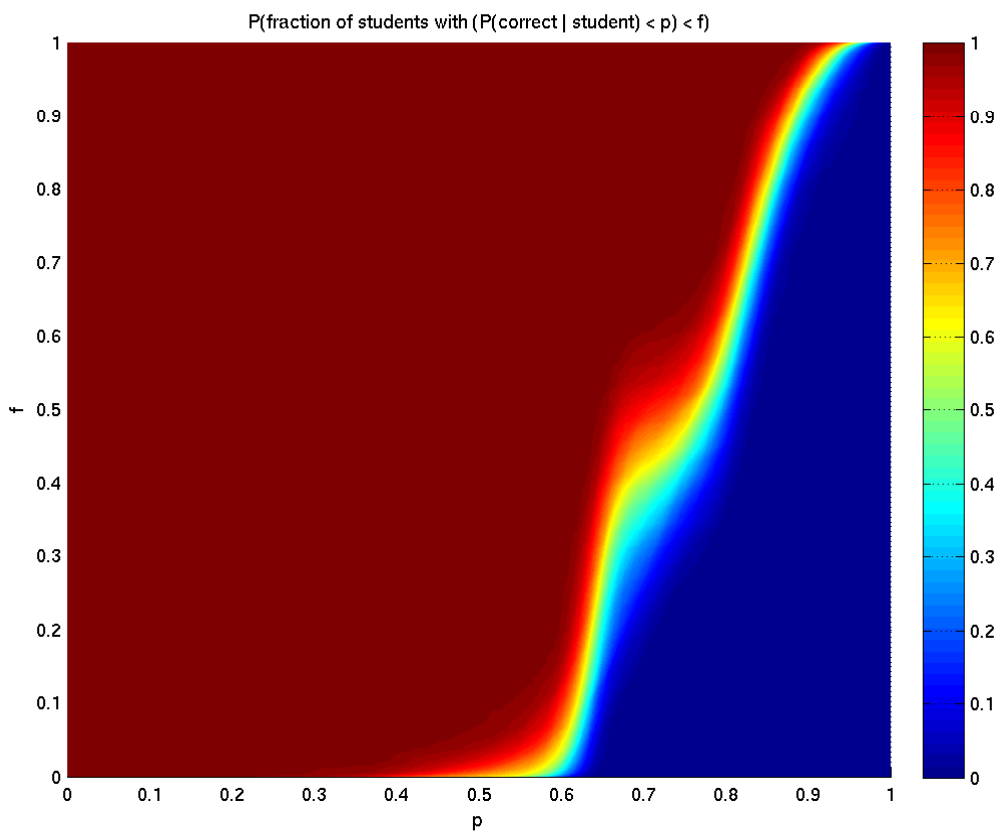} &
\includegraphics[scale=0.5]{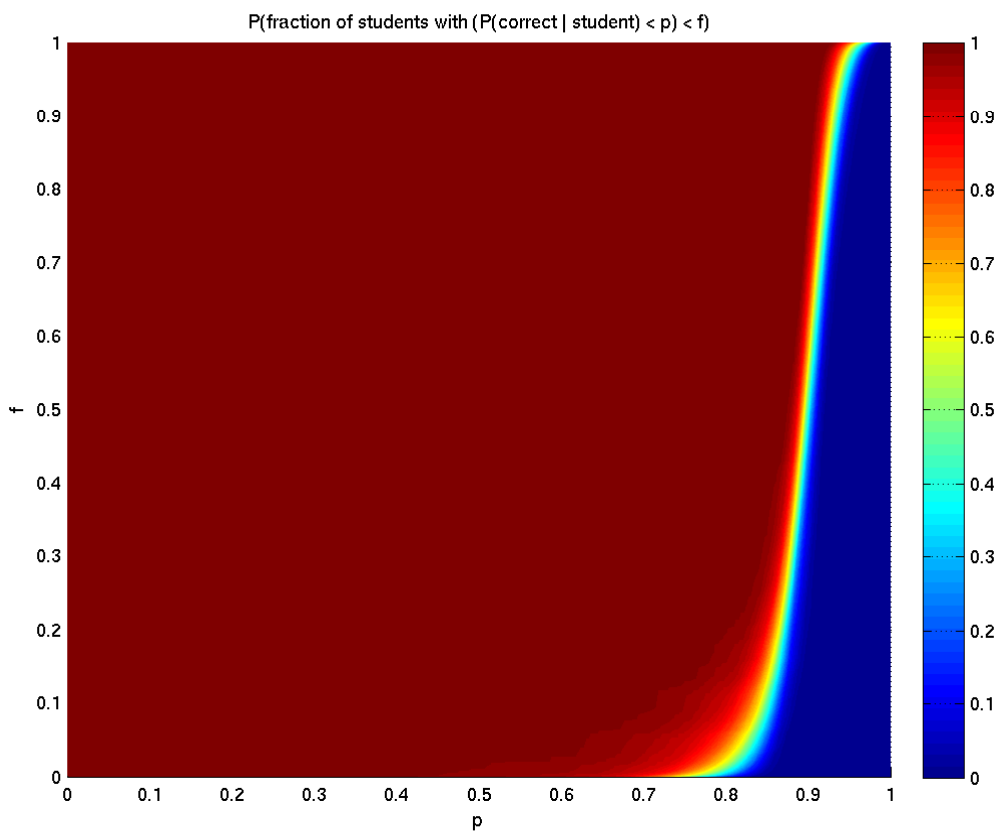} \\

\end{tabular}

\caption{Posterior distribution of the cumulative distribution
  function over students in each class of the probability of getting a
  question correct: school 4. The left two plots are for the pretest
  and the right two plots for the posttest. The top two plots are
  without SAC and the bottom two with SAC.
\label{school4cdf}
}
\end{figure}

We can also plot the cumulative distribution over students in each
class of the probability of getting a question correct together with
its posterior uncertainty. This is shown for school 2 in figure
\ref{school2cdf} and for school 4 in figure \ref{school4cdf}. These
plots are unsurprising given the plots of individual gains under the
two methods, but they do show various other differences between the
schools, such as the difference between the pretest ability of the
classes getting teaching with and without SAC (correct observation)
and the apparently greater gain in ability over the timecourse in
school 4 than school 2 (not necessarily a correct observation, as the
tests were different in each school, and school 4's posttest might
have been easier).

\subsection{Discussion}

The differences in the apparent effects of incorporating SAC into
teaching in the different schools are intriguing. 

We should first note that we are not claiming frequentist significance
for any of these results -- indeed we are not interested in
frequentist properties over many datasets as we have only one dataset
to hand. Similarly the usual frequentist corrections for multiple
analyses do not figure in this Bayesian situation.

As to the reasons for the differences, we can only speculate, while
noting that different schools and different classes have both
different students and potentially different teachers. 

For example, it may be the case that the effect of using SAC is very
dependent on the teacher's understanding of it, or indeed on whether
the teacher is a willing or unwilling user of it. Alternatively, it
might be that some teachers unconsciously spend more time with those
at the bottom of the class, or on the other hand with those at the
top. A further possibility, impossible to assess as neither sex of
teachers nor sex of students was available in the public dataset, is
that it makes a difference whether teacher or student is male or
female\footnote{RFS's wife, herself an academic with far more
publications involving statistics than her husband, seems to be
totally unable to express her own confidence levels as other than 0 or
1, thinking that she either knows something or that she doesn't; her
husband has no such difficulty. While this example does not generalise
to all men and all women, as James Damore so aptly observed in
\cite{JamesDamore}, men and women are different.}.

In future research it is highly desirable that such issues do not
arise. One obvious approach is to take enough students to fit into two
classes, randomise the students to one class or the other, and to have
the same teacher teach both classes, but one with and one without SAC;
then repeat this setup across multiple schools. Ideally public data
should include both sex of teacher and sex of each student. Something
that schools, parents, and students seem, however, to be slow to grasp
is that when it is not known whether a particular teaching method
works better than another or not, it is entirely ethical to randomise
students between the two methods -- indeed it could be argued that it
would be unethical \textit{not} to do such research, as we would then
never find out. The fact that of 55 schools approached by Foster in
\cite{FosterCA} only 4 agreed to take part is worrying for the future
of educational research in the UK.

\clearpage

\bibliography{ms}
\bibliographystyle{ieeetr}

\end{document}